\documentclass[fleqn,usenatbib]{mnras}

\usepackage{newtxtext,newtxmath}

\usepackage[T1]{fontenc}

\DeclareRobustCommand{\VAN}[3]{#2}
\let\VANthebibliography\thebibliography
\def\thebibliography{\DeclareRobustCommand{\VAN}[3]{##3}\VANthebibliography}

\usepackage{graphicx}	
\usepackage{amsmath}	

\title[Carbonates and ices in galaxy molecular clouds]{Carbonates and ices in the $\boldmath{z=0.89}$ galaxy-absorber towards PKS~1830--211 and within star-forming regions of the Milky Way}

\author[J. E. Bowey]{
Janet E. Bowey$^{1}$\thanks{E-mail: boweyj@cardiff.ac.uk}
\\
$^1$ School of Physics and Astronomy, Cardiff University, Queens Buildings, The Parade, Cardiff CF24 3AA, UK.
}

\date{Accepted XXX. Received YYY; in original form ZZZ}

\pubyear{2023}

\begin{document}
\label{firstpage}
\pagerange{\pageref{firstpage}--\pageref{lastpage}}
\maketitle

\begin{abstract}
A pair of 6.0 and 6.9~\micron\ absorption features are frequently
observed in Milky-Way (MW) molecular-clouds and YSOs; they also occur
in the $z=0.886$ rest-frame of a molecule-rich spiral galaxy obscuring
blazar PKS~1830--211. I calibrate $\chi^2$-fitting methods which match
observations with two or three laboratory spectra. The
6.0-\micron\ component is dominated by H$_2$O ice, as
expected. Included MW sources were selected using opacity criteria
which limit the range of explored H$_2$O-ice column densities to
1.6--$2.4 \times 10^{18}$ molecules~cm$^{-2}$, while the H$_2$O-ice
density in the galaxy absorber is $(2.7\pm 0.5)\times 10^{18}$
molecules cm$^{-2}$. CH$_3$OH ice and / or small (< 0.1-\micron-sized)
Ca- and Mg-bearing carbonates contribute at 6.9~\micron. The 41~\%
CH$_3$OH : H$_2$O molecular ratio in the PKS~1830--211 absorber is significantly
higher than in the molecular cloud towards Taurus-Elias 16 (<7.5\%)
and similar to the highest value in MW YSOs (35\% in AFGL~989). Fitted
carbonate (-CO$_3$) : H$_2$O ratios in the galaxy absorber of 0.091\%
are low in comparison to most of the ratios detected in the MW sample
(0.2--0.4~\%; $\sim 0$~\% in AFGL~989). Inorganic carbonates could
explain the increased oxygen depletion at the
diffuse-medium-to-molecular-cloud transition which Jones \& Ysard
associated with unobserved organic carbonates or materials with a C:O
ratio of 1:3.
\end{abstract}




\section{Introduction}
\label{sec:intro}


In 1977 \citet{puetter1977} observed Milky-Way (MW) protostars OMC
2-IRS3, GL~989, GL~2591, GL~2884 and NGC~7538 with the Kuiper Airborne
Observatory (KAO) and found absorption features near 6.0 and
6.9~\micron\ in addition to the known 3.0 \micron\ `ice' and
9.7~\micron\ `silicate' features. \citet{puetter79} subsequently
observed massive star-forming region W51--IRS2 and suggested that the
water of hydration in silicates could carry the 6.0~\micron-band,
carbonates could carry the 6.9~\micron\ feature and that hydrocarbons
could be additional components. May 1978 KAO observations of W33A
\citep{soifer1979} provided the best early data set. Some forty-five
years after discovery, the bulk of the 6.0~\micron\ band is associated
with H$_2$O ice and the origins of the 6.9~\micron\ band are
`enigmatic' with at least two components
\citep[e.g.][]{Boogert2008}. This `W'-shaped feature \citep{Aller2012}
is also seen in the $z=0.886$ face-on spiral galaxy lens obscuring
blazar PKS~1830--211 \citep{Winn2002}.

The goals of this work are to determine the primary dust component(s)
responsible for the 6.9~\micron\ band. I compare results in the
different MW and galaxy-absorber sightlines because differences in the
dust components are a consequence of chemical and physical variations
in their host environments. I use the smallest number of laboratory
spectra in order to allow the statistics of the fits
(reduced-chisquared values, $\chi^2_\nu$), to distinguish between
models. In cases where the modelling is inconclusive, I constrain the
results with information from other observations. Contributing dust
components are: H$_2$O and CH$_3$OH ices, carbonates (calcite,
dolomite, or magnesite), SiC, OCN$^-$ and CO$_2$ ice. For the
remainder of this paper the reader should assume that H$_2$O, CO$_2$
and CH$_3$OH are ices unless stated otherwise.

Typical infrared absorption bands in MW star-forming environments are
described in Section~\ref{sec:irbands}. MW source selection criteria
are explained in Section~\ref{sec:MWsel} and their characteristics are
decribed in Section~\ref{sec:sightlines}. H$_2$O, methanol (CH$_3$OH),
and carbonate models of the 6--8~\micron\ spectra are outlined in
Section~\ref{sec:6-8mod}; the laboratory data for modelling the
astronomical bands are listed in Section~\ref{sec:lab} and
Table~\ref{table}. 20-\micron-sized SiC grains are added to the
6--8-\micron-model of Mon~R2~IRS~3 in
Section~\ref{sec:sic}. Abundances of H$_2$O, CH$_3$OH, carbonates and
SiC are derived in Section~\ref{sec:abundances}. In
Section~\ref{sec:2.5-5mod}, shorter-wavelength bands at 4.3 and
4.6~\micron\ in the PKS~1830--211 spectrum are associated with CO$_2$
and OCN$^-$, respectively. The quantitative results are summarised in
Section~\ref{sec:summres} and the observational and theoretical
consequences of a population of carbonate dust discussed in
Section~\ref{sec:impact}. The conclusions are in
Section~\ref{sec:conclusions}.



\section{Infrared Absorption bands in MW molecular-clouds and YSOs}
\label{sec:irbands}
3--12~\micron~ spectra of YSOs and sightlines through molecular clouds
towards background stars within the MW contain several features
between the 3.0~\micron\ H$_2$O `ice' and 9.7~\micron\ `silicate'
bands. They include:

\subsection{Bands between 3 and 5~\micron}
Absorption bands near 3.25, 3.44 and 3.47~\micron\ are not a part of
this study because they do not appear in the PKS~1830--211
galaxy-absorber spectrum; they have recently been modelled with PAHs
and hydrogenated PAHs co-condensed with H$_2$O ice \citep{Chiar2021}.
While the 3.25~\micron\ band is normally associated with PAHs the
3.47~\micron\ band is typically attributed to nano-diamonds
\citep[e.g.][]{allamandola1992}. Bands at 3.54 and 3.95 are used to
obtain CH$_3$OH abundances \citep[e.g.][]{Gibb2004}. Two, normally
prominent, narrow absorption bands at 4.27~\micron\ (fwhm $\approx
0.03$~\micron) and 4.61~\micron\ (fwhm $\approx 0.06$~\micron) are
assigned to the C=O stretch in pure CO$_2$ or mixed CO$_2$ bearing
ices, and a CN stretch in OCN$^{-}$ or XCN, respectively
\citep[e.g.][]{Gibb2004, Brucato2006}.

\subsection{The 6.0~\micron\ H$_2$O and carbonaceous band}
The non-gaussian 6.0~\micron\ (fwhm$\approx
0.5$~\micron) band associated with H$_2$O ice is thought to contain
additional contributions from carbonaceous materials. Possible
contributors include: Organic Refractory Material 
\citep[ORM;][]{GibbWhittet2002}, a 5.85~\micron\ C$=$O (carbonyl) band
which occurs in Hydrogenated Amorphous Carbon (HAC) formed in an
oxygen-bearing atmosphere \citep{Grishko2002}, and formic acid (HCOOH)
as well as formate (HCOO$^-$) ion which might contribute at
6.33~\micron\ \citep[see][]{Boogert2008}.

\subsection{An `enigmatic' band near 6.9~\micron}

\subsubsection{CH$_3$OH ice}
CH$_3$OH \citep[e.g.][]{Gibb2004} is a reasonable match to the $\sim
0.4$~\micron-wide 6.9~\micron\ band in many environments but the
inferred abundance is frequently higher than estimates derived from
the CH$_3$OH bands at 3.54~\micron\ and 3.95~\micron. In addition, the
peak-wavelength of the astronomical band shifts between sources whilst
the CH$_3$OH peak does not, so two components are thought to
contribute \citep{Keane2001,Boogert2008,Boogert2011}.

\subsubsection{Aromatic and aliphatic candidates}
Combinations of Polycyclic aromatic hydrocarbons (PAHs) have been
found too narrow to fit the features in star-forming regions
\citep{Chiar2021,Mattioda2020}. However, a narrower
6.9~\micron\ absorption feature in the ISO spectrum of the line of
sight towards Sgr A* (the MW Galactic Centre) which includes
molecular-cloud and diffuse medium dust, is associated with asymmetric
CH deformation modes in diffuse-medium aliphatic hydrocarbons
\citep{Chiar2000} because it is insufficiently broad to match
CH$_3$OH. This narrower feature has also been identified in the
ice-free diffuse medium sightline towards Cyg~OB2~no.12
\citep{Hensley2020}.

\subsubsection{Oxygen-rich candidates--silicates and carbonates}
Inorganic carbonates (of mineralogical formula X--CO$_3$, where X is
usually a combination including Mg, Ca, or Fe) were early candidates
for the 6.9~\micron\ band \citep{SA1985} due the occurrance of a
strong band \citep[arising from an asymetric stretch within the
  CO$_3^{2-}$ ion; see][]{White1974} near to this wavelength.
However, \citet{Keane2001} excluded them from consideration because
the 0.6~\micron-wide carbonate bands in existing laboratory spectra
were broader than the astronomical features. \citet{BH2005} found a
match with the overtone spectrum of a crystalline silicate from the
melilite (Ca$_2$Mg(Si,Al)$_2$O$_7$) group, but laboratory measurements
of a set of melilites \citep{BH2022} were used to prove that the
`melilite' band was produced by minor (<0.1 \% by mass) contamination
of the sample with carbonate powder \footnote{True melilite overtones
were a pair of narrow (<0.1~\micron-) peaks at 6.4~and
6.8~\micron.}. With the initial purpose of correcting the `melilite'
spectra, \citet{BH2022} obtained spectra of very thin $\sim
0.04$--0.15~\micron-thick carbonate films: the 6.9~\micron\ absorption
bands of these materials were narrower than those in the
KBr-dispersion\footnote{\label{shapensize} Transmission spectra
derived from dispersions of powder mixed with other matrix material
(KBr, polyethylene) have broader spectral peaks than those of thin
films due to the effect of scattering between the grains and the
matrix. \citet{BHK2020} contains a detailed discussion of experimental
methods and comparison between KBr and thin-film spectra for
pyroxenes.  Band-widths also increase if the grain-sizes or
sample-thicknesses or grain sizes are too large because light is not
transmitted at the band centre; Hofmeister's preliminary spectra of
thicker 0.2--0.3~\micron\ carbonate films were significantly broader
than the data in \citet{BH2022} and do not match astronomical
spectra. Spectral artefacts are discussed in detail by
\cite{HKS2003}. In the astronomical literature from the 1980s to 2000s
this phenomenon is frequently referred to as `grain-shape and size
effects' and is one of the reasons why the use of particulate-spectra
was frowned on in astronomy.} spectra used by \citeauthor{Keane2001}
and provided a good match to a similar 6.9~\micron\ absorption band
produced by dust obscuring the carbon-rich atmosphere of Sakurai's
Object (V4334 Sgr) in 2005--2008 Spitzer observations. Interpretation
of Sakurai's Object spectra \citep{Bowey2021,BH2022} was simplified by
the absence of ices as exemplified by the absence of a
6.0~\micron\ H$_2$O-ice absorption band. PAH absorption was added to
match finer structure in its 6--7~\micron\ spectra and fits improved
if overtone features due to large (25~\micron-sized) SiC grains were
included in the model.

Here, I shall use the same carbonate spectra to model the sources in
environments where the fitting is complicated by the co-existence of
ices in the same lines of sight, and add SiC grains when
necessary. The contribution of PAHs will be briefly considered, but is
not a focus of this research.

An absence of experimental studies of carbonate formation under
astronomical conditions, is frustrating, but unsurprising (see
\citet{BH2022} for a detailed discussion). However, carbonates are
candidate carriers of a broad 90-\micron\ emission band in the spectra
of planetary nebulae (PNe) \citep[e.g][]{Kemper2002} and YSOs
\citep[e.g][]{Ceccarelli2002}. Carbonates were also included in models
of the spectra of freshly produced dust from the deep impact
experiment on comet Tempel~1~\citep{Lisse2007} because they produce a
6.9~\micron\ absorption band in the spectra of pyroxene and
layer-lattice interplanetary dust particles~\citep{SA1985}. They are
found with hydrated minerals in meteorites
\citep[e.g.][]{RubinMa2017}, and form rapidly from CaO exposed to the
air under ambient terrestrial conditions \citep{GR2009}.

\section{Selection of MW Sources}
\label{sec:MWsel}

Due to the overlapping wavelengths of the CH$_3$OH and carbonate
6.9-\micron\ bands, fits to the 5--8~\micron-spectra were compared
with CH$_3$OH to H$_2$O abundance-ratios from 3--4~\micron\ data for the
MW targets. Hence, MW spectra were selected according to the following
criteria:
\begin{enumerate}
\item There must be good spectra of the 6--8~\micron\ absorption bands
  which are unsaturated, i.e. the absorption peaks are curved rather than
  flat-topped. Laboratory spectra of rounded and flattened peaks
  indicate that the grains are too large or the sample is too thick
  for light to pass through to give a reliable spectral shape \citep[e.g][]{HKS2003} and footnote~\ref{shapensize}.
\item There must be good unsaturated data for the 3.0~\micron\ H$_2$O
      band and estimates, or upper limits of the CH$_3$OH abundance
  from the 3.54~\micron\ and 3.95~\micron\ bands.
\end{enumerate}

The sample includes the quiescent sightline through the Taurus
molecular cloud towards Taurus Elias~16 and spectra of YSOs in
high-mass star forming regions (S140~ IRS~1, Mon~R2~IRS~3, AFGL~989).
Unfortunately, these criteria eliminated observations of low-mass YSOs
like $\rho$-Elias~29 (the 6--8~\micron\ spectrum was weak and
relatively noisy), well-known high-mass YSOs like W~33A (saturated
3.0~\micron\ band) which have more ice features than discussed here,
and quiescent-molecular cloud sources (no 3.0~\micron\ spectra), and
sightlines where the 6--8~\micron\ absorption bands were coincident
with PAH emission. The 3--4~\micron\ saturation constraint limits the
range of explored H$_2$O ice column densities and three of the four
3--4~\micron\ CH$_3$OH estimates are upper limits.

\section{Sightline Characteristics}
\label{sec:sightlines}
\subsection{PKS~1830--211}

The line of sight towards blazar PKS~1830--211 at $z=2.507$ is
physically and chemically interesting at radio, submm and IR
wavelengths because its light passes through a face-on lensing spiral
galaxy at a redshift of $z=0.886$ \citep{Winn2002}. Two or three
lensed components are observed in ALMA images with a maximum
separation of 1~\arcsec \citep{Muller2020}. The SW component of
PKS~1830--211 is currently the only high redshift object in which
gas-phase methanol (CH$_3$OH) absorption has been detected in the
submm \citep{Muller2021}; this together with other submm and radio
absorption bands due to 60 gas-phase molecular species indicate
conditions within the galaxy-absorber are similar to those in
Milky-Way (MW) cold molecular clouds and hot UV-rich HII regions
associated with young stellar objects (YSOs)
\citep{Tercero2020,Muller2021}.

PKS~1830--211 was a single object in the $3.7\arcsec \times
57$\arcsec\ aperture of the Short-Low (SL) module of the infrared
spectrometer (IRS) on Spitzer; its spectrum is known for its unusual
10~\micron\ silicate absorption feature which resembles crystalline
olivine \citep{Aller2012}, rather than the glassy silicates common to
other galaxies and the interstellar medium of the MW; early MIRI/JWST
observations are scheduled \citep{AllerJWST2021}. \citet{Aller2012}
also remarked on the similarity between the galaxy-absorber's
additional 6.0~\micron\ and 6.9~\micron\ bands and those in MW
embedded ice-rich YSOs and associated the 6.0~\micron-band with H$_2$O
ice. The data selection process for this work is explained in
Appendix~\ref{app:PKSobs}.



\subsection{MW Dark molecular cloud: Taurus Elias~16}
The interstellar line of sight towards the field-star Elias~16 is the
archetypal quiescent dark and dusty molecular cloud environment
because it is a bright and highly reddened field star
(E(J-K$_s$)=4.76; A$_V \sim$19) of spectral type K1 III behind the
Taurus Molecular Cloud \citep{Chiar2007}. Spectral studies of ices,
PAHs and silicates, and spectropolarimetric studies of ices have all
been made towards Elias~16
\citep[e.g.][respectively]{Knez2005,Chiar2021,Bowey1998,Hough2008}.
Spitzer data in this work are from the CASSIS archive; the original
observations were published by \cite{Knez2005}. Short Wavelength
Spectrometer \citep{deGraauw1996} Infrared Space Observatory (ISO)
data published by \citet{Gibb2004} were used for analysis in the
2--5-\micron\ range.

\subsection{MW Massive Star-Forming Regions}
SWS/ISO spectra published by \citet{Gibb2004} were used for S140~IRS~1, 
AFGL~989 and Mon~R2~IRS~3.

S140 is an H II region located 910 pc away behind $\sim23$ mag of
extinction in a molecular cloud, which is forming high- and low-mass
stars~\citep{Evans1989}. K'~band images \citep{PreibischSmith2002} show
several high mass YSOs (8--10 M\sun) within the 30--40\arcsec\ ISO
beam.  S140~IRS~1 has a dust disc which has been resolved at 1.3mm
\citep{MaudHoare2013}.

AFGL~989, otherwise known as Allen's Source, is the brightest IR
source in NGC~2264 (IRS~1). It is a high-mass YSO (10~M\sun) which is
invisible in the optical. Mid-infrared interferometric observations
indicate that the object is surrounded by a flat circumstellar disk
that has properties similar to disks typically found around lower-mass
young stellar objects \citep{Grellmann2011}. The luminosity is
consistent with a 9.5 M\sun\ B2 zero-age main-sequence star
\citep{Allen1972} with 20--30 mag of visual extinction
\citep{Thompson1998}.

In the near-infrared, Mon R2 IRS 3 is a bright 500-au conical
reflection nebula containing two to three massive early-type stars
(IRS 3N, IRS 3S and possibly IRS 3 NE). Infrared speckle imaging
suggests that the conical-shaped nebula is due to collimation of the
light of IRS 3 by a 500-au disc \citep{Koresko1993} and the presence
of three further sources within 2.6 arcsec of IRS 3S; the $A_V$ to the
primary is >30~mag \citep{Preibischetal2002}.


\section{Ice and carbonate models of 6--8~\micron\ spectra}
\label{sec:6-8mod}
The infrared continua of objects with strong absorption features are
poorly constrained longward of
5~\micron\ \citep[e.g. see][]{Gibb2004,Boogert2008} and it is common
practice to use a low-order polynomial or spline fit over as wide a
wavelength range as possible and then to add the absorbers for each of
the bands. In contrast to this approach, I select laboratory data for
thin films of candidate absorbers and simultaneously fit the optical
depths and continua. My continua are represented by simple
mathematical formulae over the narrower wavelength range of the
absorption features; these represent an ill-defined combinination of
the effects of baseline subtraction from the laboratory spectra, 
source physics, and foreground extinction.

Since ice abundances are usually given in terms of number of molecular
absorbers (i.e. molecular density) whilst abundances of the refractory
components (i.e. the grain cores) are quoted as mass and grain number
densities, I give estimates of all three parameters. The relationship
between these quantities is described in Appendix~\ref{app:abuncalc}.

\begin{figure*}
\includegraphics[bb= 0 450 500 660, height=0.17\textheight]{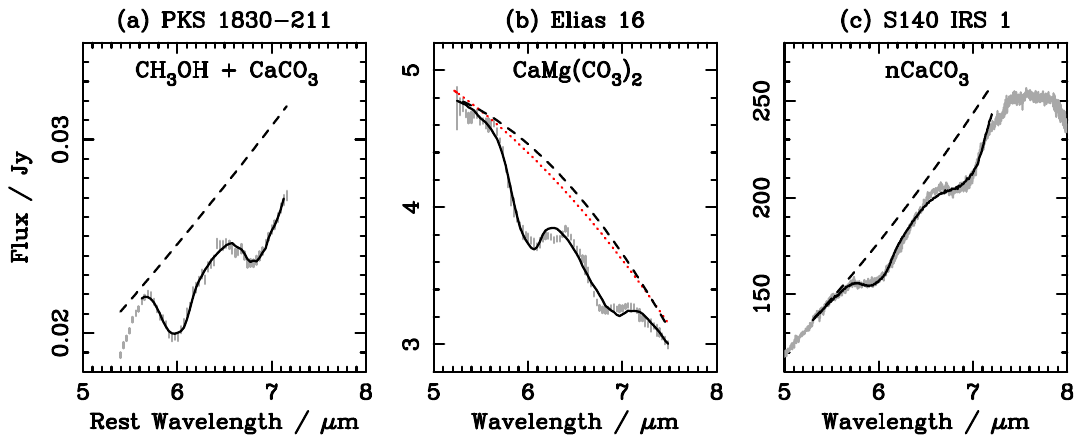}
\includegraphics[bb= 0 450 390 660, height=0.17\textheight]{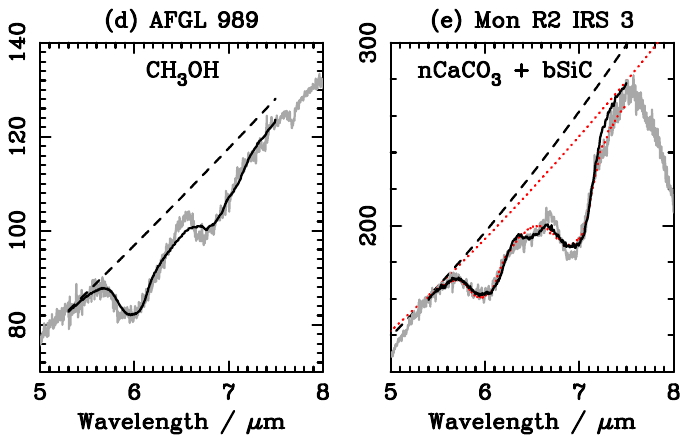}
\caption{Best fits (solid black) to absorption features in the 6--8~\micron\ spectra of
  of PKS~1830--211 and MW sources: Taurus~Elias~16, S140~IRS~1, AFGL~989 and Mon~R2~IRS~3 (grey; error bars are plotted on the Spitzer data, but not the ISO data) with 10-Kelvin H$_2$O and a carbonate or CH$_3$OH or both 6.9~\micron\ components. Dashed curves are continua derived with the best models. Only Mon~R2~IRS~3 required a bSiC component. Red dotted curves indicate the continuum derived for Elias~16 with the poor two-component CH$_3$OH model and the continuum and fit to Mon~R2~IRS~3 if the bSiC component were excluded.
  \label{fig:natfits}}
\end{figure*}
\begin{figure*}
\includegraphics[bb= 0 35 500 760, height=0.6\textheight]{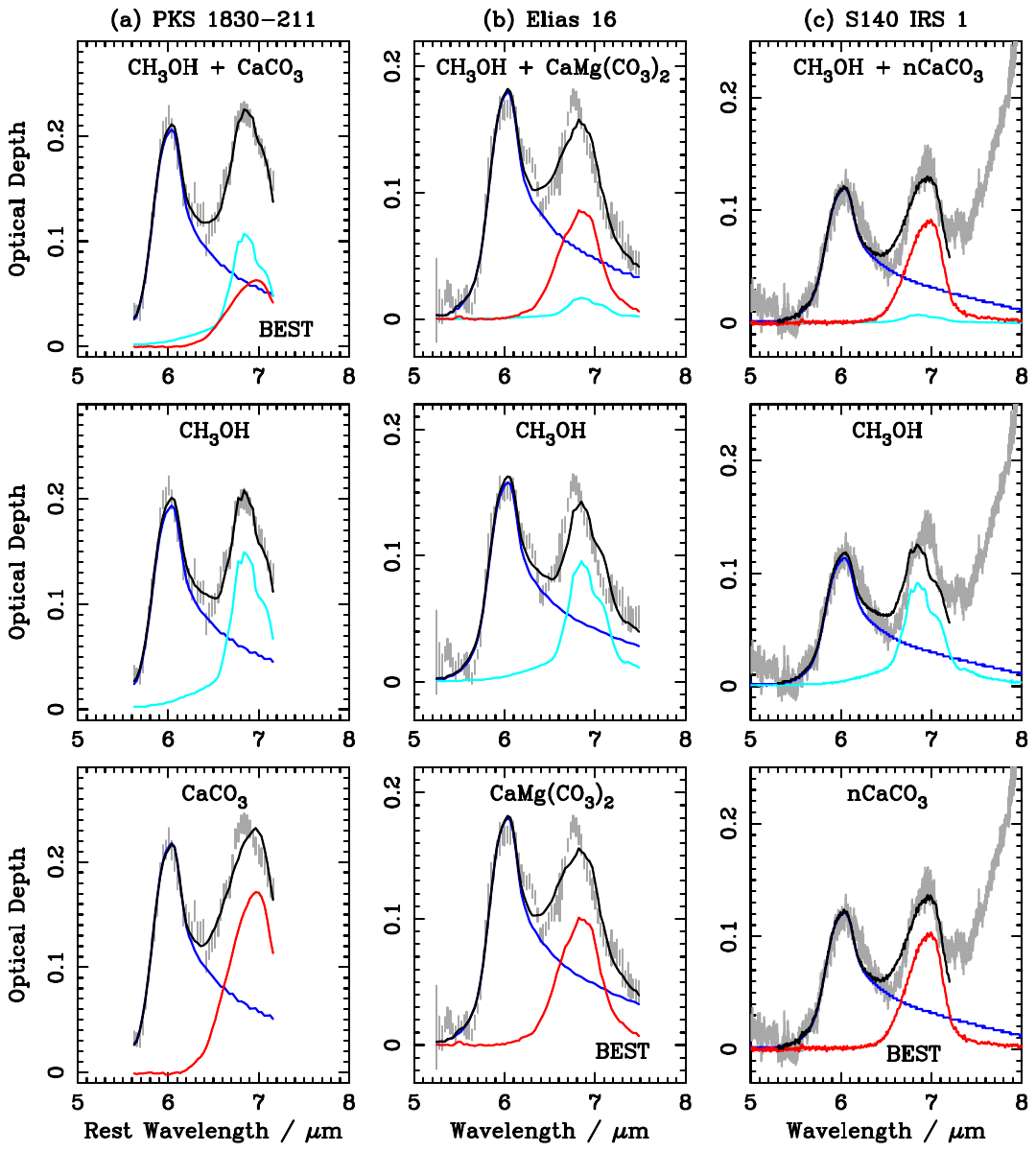}\includegraphics[bb= 0 35 390 760, height=0.6\textheight]{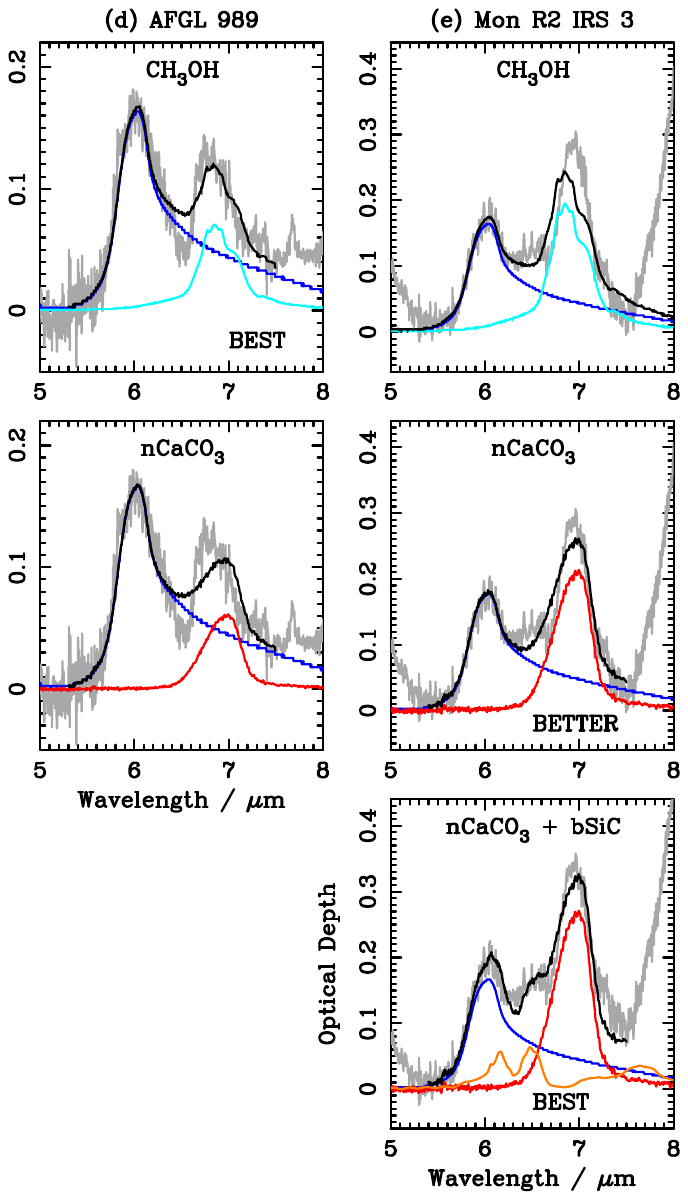}\\

\caption{5--8~\micron\ optical depth spectra (grey) with combinations
  including 10-Kelvin H$_2$O (blue), a carbonate (red) and/or CH$_3$OH
  (cyan). Black curves are the sum of the components over the fitted
  range. The `BEST' Mon~R2~IRS~3 fit included an additional bSiC
  component (orange) to fit a plateau at 6.4~\micron. Three-component
  fits to AFGL~989 and Mon~R2~IRS~3 are not included because they are
  indistinguishable from the `BEST' (AFGL~989), or `BETTER'
  (Mon~R2~IRS~3) two-component fits. \label{fig:natopt}}
\end{figure*}

\subsection{Laboratory spectra used to model the absorption bands}
\label{sec:lab}
Each continuum function, $C_\nu$, was extinguished by up to three
components represented by H$_2$O and CH$_3$OH, and one of the carbonate spectra listed in Table~\ref{tab:const}.
\begin{equation}
\label{eq:ext}  
  F_\nu=C_\nu\exp{(-\sum^3_{i=1}  c_i\uptau_i(\lambda))}, 
\end{equation}
where $\uptau_i$ is the shape of the $i^{th}$ absorber,
normalised to unity at the tallest peak in the wavelength range of
interest. 
\begin{table*}
\begin{minipage}{\linewidth}
\caption{Laboratory data: peak wavelength, $\lambda_{pk}$, sample thickness (grain size), $d$, mass density ($\rho$), mass absorption coefficient, $\kappa_{pk}$ at the peak wavelength, peak width, $\Delta_\nu$, integrated band strength, $A_i$ and the frequency range for the $A_i$ calculation. $\kappa_{pk}$ and $A_i$ are derived in Appendices~\ref{app:mac} and \ref{app:molabs}, respectively. \label{tab:const}}
    \begin{tabular}{lllllllll}
\hline
      Absorber&$\lambda_{pk}$
      &$d$\footnote{Sample thickness, $d$, is used as a proxy for grain length in refractory components (see Appendix~\ref{app:grainden} for justification). The nominal volume is $d^3$.}&$\rho$&$\kappa_{pk}$&$\Delta_\nu$\footnote{fwhm}&$A_i$&$\nu_1$ --$\nu_2$\\
      &\micron&\micron&gcm$^{-3}$&cm$^2$g$^{-1}$&cm$^{-1}$&cm molecule$^{-1}$&cm$^{-1}$&Ref.\footnote{\label{foot:ref} References: Spectrum and film thickness from 1-- \citet{Hudgins1993}, 2-- \citet{BH2022}, 3--\citet{Hof2009}; spectrum from 4 --\citet{vanBroekhuizen2006}, 5--\citet{Brucato2006}. References for ice mass densities, $\rho$ : 6-- H$_2$O at 20K --\citet{Dohnalek2003}; 7-- CH$_3$OH at 20~K--\citet{Luna2018}; 8-- CO$_2$ at 80~K \citet{Mangan2017}. Refractory measurements and densities were obtained at room temperature.}\\
\hline
\multicolumn{9}{c}{6--7~\micron\ components}\\
\multicolumn{9}{l}{{\bf Ices}}\\
H$_2$O & 6.05&0.41        &0.94&1700        &144&$1.1\times 10^{-17}$&1100--1900&1, 6 \\
      CH$_3$OH& 6.85&0.56&0.64&3000                &115&$1.1\times 10^{-17}$&1200--1800&1, 7\\
\multicolumn{9}{l}{{\bf Carbonates}}\\
      Dolomite, CaMg(CO$_3$)$_2$& 6.87&0.15&2.86&33000&120 &$1.94\times 10^{-15}$\footnote{To follow chemical convention, the formula has two CO$_3$ groups because Fe and Mg are interchangeable in the same lattice positions. Quoted band strength is the value per CO$_3$ group to match the other carbonates.}&1200--1700&2\\
      Magnesite, MgCO$_3$\footnote{Magnesite provided a good fit to data for Sakurai's Object \citep{BH2022}, the integrated cross-section and fwhm is included for completeness.}&6.87&0.15&2.98&27000&120&$1.53\times 10^{-15}$&1200--1700&2\\
      Calcite, CaCO$_3$& 6.97&0.11&2.72&49000       &125        &$3.24\times 10^{-15}$&1100--1640&2\\
      Calcite, nCaCO$_3$\footnote{The prefixed n in the chemical formula is to distinguish the very thin film.}& 6.97&$\sim 0.04$ &2.72&63000     &96          &$3.21\times 10^{-15}$&1100-1640&2\\
\multicolumn{9}{l}{\bf Silicon Carbide overtone spectrum}\\
      bSiC \footnote{Overtone bands in 6H Alfa/Aesar sample orientated $E \perp c$; Sakurai's Object was better fitted with a different sample (a 25~\micron-thick $\beta$~SiC wafer).} &6.48&22&3.2&210&55&$2.4\times10^{-18}$\footnote{blended with peak at 6.17~\micron}&1460--1585&3\\
    \\
\multicolumn{9}{c}{4--5~\micron\ components}\\
\multicolumn{9}{l}{{\bf Ice}}\\

$\nu_3$ CO$_2$&4.27&&1.68&34000&18&$7.6\times 10^{-17}$\footnote{pure CO$_2$ band strength \citet{Gerakines1995}}&-&4, 8\\
\multicolumn{9}{l}{{\bf Organic Residue}}\\

      OCN$^{-}$ &4.61       &   &    &     &$\approx 30$\footnote{estimated from \citet{Brucato2006} data}       &$1.3\times 10^{-16}$ \footnote{\citet{vanBroekhuizen2004}, accuracy $\pm 20$\%}&-&5\\

      \hline
    \end{tabular}
  \end{minipage}
\end{table*}

Three combinations of materials were fitted to each source spectrum:
(i) H$_2$O, CH$_3$OH and a carbonate, (ii) H$_2$O and CH$_3$OH, (iii)
H$_2$O and a carbonate. The ice spectra \citep{Hudgins1993} were
measured at 10~Kelvin, the carbonate spectra \citep{BH2022} were
obtained at room temperature (there are no suitable low-temperature
data). The carbonate samples included magnesite (MgCO$_3$), dolomite
(CaMg(CO$_3$)$_2$) and calcite (CaCO$_3$). Peak wavelengths vary
between 6.87~\micron\ (magnesite, dolomite) and
6.97~\micron\ (calcite); most of the spectra were obtained from 0.1-
to 0.15-\micron-thick powder films. The effect of grain-size on the
calcite fits was explored by using the spectrum of an additional $\sim
0.04-$\micron-thick calcite film (nCaCO$_3$). Only the best carbonate
fits are included in the paper; fit qualities are governed primarily
by the match to peak-wavelength and secondly by feature-width. Small
components of other materials may also contribute to the spectra,
especially to the 6.0-\micron\ H$_2$O band.\footnote{Organic residues
may broaden and deepen the feature \citep{GibbWhittet2002} but were
included by them to explain extra depth in the 6.0\micron\ band in
comparison to predictions from the 3.0-\micron\ H$_2$O band. I did
not included an (organic refractory material) ORM component because the
fits were degenerate.} In tests extra components fell to zero or the fit
was unconstrained.

\subsection{Continua and fitting process}
\subsubsection{Power-law galaxy and YSO continua}
The observed spectra of PKS~1830--211, and the YSOs,
were modelled by inserting a wavelength-dependent power-law  continuum,
\begin{equation}
\label{eq:pow}
C_{\nu}=c_0\lambda^{\alpha},
\end{equation}
into equation ~\ref{eq:ext}. Constant, $c_o$, and optical-depth scale
factors $c_1$ to $c_3$ were constrained to positive values (to prevent
the introduction of spurious emission features) and optimized by the
downhill-simplex method of $\chi^2$-minimization as implemented by the
AMOEBA routine in Interactive Data Language (IDL). Power-law index,
$\alpha$, values were selected by trial and error to give the best fit
to each source.

\subsubsection{Taurus-Elias~16 continuum}
Power-law models which match the featureless Elias~16 spectrum between
5.0 and 5.6~\micron\ and the 6.0~\micron\ band are too shallow to
match the 6.9~\micron\ band and continuum beyond 7.0~\micron\;
$\chi^2$ values are poor. Unlike the other sources which are
characterised by largely optically-thick power-law emission local to
the source, Elias~16 is a heavily reddened field star which is better
represented by a second order polynomial in
wavelength:

\begin{equation}
\label{eq:pol}
C_\nu=a_0+a_1\lambda+a_2\lambda^2.
\end{equation}
in equation~\ref{eq:ext}

Six parameter fits of coefficients $a_0$ to
$a_2$ and optical-depth scale factors $c_1$ to $c_3$ were poorly
constrained due to the excessive number of degrees of freedom; five
parameter fits were constrained.  Therefore, values of the
6.0~\micron\ H$_2$O optical depth were pre-selected by trial-and-error
to give the lowest $\chi^2_\nu$ in the H$_2$O, CH$_3$OH and carbonate
models.  H$_2$O components in two-component models including either
CH$_3$OH, or a carbonate were allowed to vary freely.

Unsophisticated tests of power-law reddened Spitzer library spectra of
KIII giants produced continua similar to the polynomial models over
this wavelength range.

\subsubsection{Calculation of fit uncertainties}
\label{sec:sigma}
Fit robustness was checked by determining one-sigma confidence
intervals after convergence at $\chi^2(fit)$. Each parameter was
shifted from the solved value by a few per cent and a new
$\chi^2(shifted)$ calculated. Then the AMOEBA routine was invoked to
minimize $| \chi^2 (shifted) - \chi^2(fit) -1.0 |$. The value at
convergence is an estimate of the one sigma confidence
interval. Formal uncertainties of well-constrained parameters are
usually small (0.1--3\%) in comparison to my estimates of systematic
uncertainties; these include the true number of dust components, the
true continuum, and uncertainties intrinsic to baseline subtraction in
the laboratory data. Hence, parameters are quoted to no more than two
significant figures despite the confidence intervals indicating
smaller uncertainties. Quoted uncertainties are based on propagating
the most pessimistic fit (5 \%) and realistic systematic (15\%)
errors.

\subsection{Fitted components and optical depth profiles}
\begin{table*}
\begin{minipage}{\linewidth}
  \caption{Fits to 6--8~\micron\ spectra with H$_2$O and CH$_3$OH and carbonates (--CO$_3$) in order of decreasing fit quality for each source. The best fits, plotted in Figure~\ref{fig:natfits}, are indicated in bold. Italics indicate optical depths predicted from 3--4~\micron\ spectra. $\sigma$-values are the one-sigma confidence intervals quoted as a percentage of each fitted optical depth to one significant figure (see Section~\ref{sec:sigma}) .\label{table}}.
\begin{tabular}{llllllllllll}
\hline
Source&Index &\multicolumn{3}{c}{H$_2$O Component}&\multicolumn{6}{l}{Additional Components}&Quality\\
&$\alpha$&$\uptau_6$(H$_2$O)&$\sigma$&$\uptau_{6.0}{(B)}$\footnote{Published estimate deduced from 3.0~\micron\ band and the temperature of the 3.0-\micron\ laboratory spectrum, see Section~\ref{sec:h2oab}. Value for Elias~16 is from \citet{Boogert2011}; others are from \citet{Boogert2008}}&&$\uptau_6$(CH$_3$OH)&$\sigma$ &$\uptau_{6.85}(P)$\footnote{Estimate defined in Section~\ref{sec:ch3ohab}. It is based on published ratios of the 3.54~\micron-CH$_3$OH ice to the 3.0~\micron-H$_2$O ice band.}&$\uptau_6$(--CO$_3$)&$\sigma$&$\chi^2_\nu$\\
\hline  
PKS~1830--211&
{\bf 1.44}&{\bf 0.21} &1&--&{\bf CH$_3$OH+CaCO$_3$}  &{\bf 0.11} &2 &&{\bf 0.063} &4 &{\bf 1.1} \\ 
&1.32      & 0.19 &1    &&   CH$_3$OH             &0.15  &2     &--& --&  &1.3 \\
&1.52      & 0.22 &1    &&   CaCO$_3$             &  --  &     &&0.17 &1&2.1  \\
 \\
Elias~16&     {\bf -- } & {\bf 0.18} &1&{\it 0.15 (10~K) }&{\bf CaMg(CO$_3$)$_2$}&{\bf --}& &&{\bf 0.10} &2&{\bf 3.52}\\
           & --       &  0.18  &--   &            &CH$_3$OH+CaMg(CO$_3$)$_2$&0.017& 10 &{\it <0.006} &0.086 &2&3.54\\
           & --       &  0.16 &2    &            &CH$_3$OH&0.096& 1&&--&&4.37\\
 \\
S140~IRS~1 &{\bf 2.07}&{\bf 0.12} &0.1&{\it 0.16 (100~K)}&{\bf nCaCO$_3$}&{\bf -- }& &&{\bf 0.10} &0.2 &{\bf 13.052}\\
       &2.05&0.12 &0.2 &&CH$_3$OH+nCaCO$_3$&0.0073& 3& {\it <0.01}&0.092& 0.3&13.055\\
       &2.04&0.11 &0.2&&CH$_3$OH&0.092&0.2&&--&&18.27\\
 \\       
AFGL~989
          &{\bf 1.25}&{\bf 0.16}& 0.2&{\it 0.18 (40~K)}&{\bf CH$_3$OH}&       {\bf 0.071} &0.6&  &{\bf --}&&{\bf 10.538}\\
&1.25&0.16 &0.2 & &CH$_3$OH+nCaCO$_3$&0.071& 0.6 & {\it 0.046} &$<10^{-6}$& &10.540\\
&1.23&0.17 &0.2& &nCaCO$_3$&--&&&0.061 &0.7 &12.44\\
 \\
Mon~R2~IRS~3&
1.65&0.18 &0.2&{\it 0.20 (100~K)}&CH$_3$OH+nCaCO$_3$& $<10^{-6}$& & {\it <0.01}&0.21 &0.2&26.9 \\
               &1.66&0.18 &0.2 &&nCaCO$_3$&--&&&0.21& 0.2 &27.0 \\
               &1.65&0.16 &0.2&&CH$_3$OH&0.20& 0.2 &&--&&42 \\
 \\
&&&&&$\uptau_{6}${(bSiC)}\\
               &{\bf 1.88}&{\bf 0.17} &0.2 &&{\bf nCaCO$_3$ + bSiC}&{\bf 0.064}& 0.6&&{\bf 0.27}& 0.1 &{\bf 20.0}\\

\hline
\end{tabular}
\end{minipage}
\end{table*}
Fits are listed in Table~\ref{table} where the best fits (with the
lowest $\chi^2_\nu$ values) are indicated in bold and shown with the
observed fluxes in Figure~\ref{fig:natfits}. Optical depth profiles
and absorption components obtained from visually distinguishable
models are shown in Figure~\ref{fig:natopt}; they were derived by
taking the natural log of the ratio of the observed, or fitted flux (equation \ref{eq:ext})
to the continuum model (equations \ref{eq:pow} or \ref{eq:pol}, as appropriate).
\begin{equation}
\uptau(\lambda)=\ln({C_\nu/F_\nu}).
\label{eq:opt}
\end{equation}


The spectrum of the PKS~1830--211 galaxy-absorber was fitted best with
a three-component model including H$_2$O, CH$_3$OH and CaCO$_3$
(Figure~\ref{fig:natfits}(a)).  Carbonate-bearing fits to Elias~16
(with CaMg(CO$_3$)$_2$; Figure~\ref{fig:natfits}(b)) and S140~IRS~1
(with nCaCO$_3$; Figure~\ref{fig:natfits}(c)) were nearly
statistically identical irrespective of the inclusion or exclusion of
CH$_3$OH. Hence, I represent the data with two-component H$_2$O and
carbonate models. In contrast, the carbonate component in AFGL~989
(Figure~\ref{fig:natfits}(e)) was negligible
($\uptau$(nCaCO$_3$)$<10^{-6}$) so the two-component H$_2$O and
CH$_3$OH model was selected.

The Mon~R2~IRS~3 observation was initially matched with a
two-component H$_2$O and nCaCO$_3$ model due to the negligible fitted
CH$_3$OH component ($\uptau$(CH$_3$OH)$<10^{-6}$) (labelled {\bf
  BETTER} in Figure~\ref{fig:natfits}(e)). However, this fit does not
match a plateau centred at 6.4~\micron\, indicating the need to add a
third absorber.

\subsection{20~\micron-sized SiC grains in Mon~R2~IRS~3}
\label{sec:sic}
Isotope measurements of thousands of 0.1--20~\micron-sized meteoritic
SiC grains \citep[e.g.][]{Hoppe1994,Speck2005} suggest that they exist
in protostellar environments. However, the SiC stretching band near
11.5~\micron\ has not been detected beyond carbon
stars\citep[e.g][]{Whittet1990}; due to the high opacity of the band
only nanometre-sized grains produce unsaturated 11.5~\micron\ features
\citep[see][]{Hof2009} so the meteoritic grains would be
opaque. Noting that larger $\la 25~\mu$m SiC grains, hereinafter
denoted bSiC (for ``big'' SiC grains), might be detectable by using
their weak overtone peaks near 6.2~ and 6.5~\micron\ \citep{Hof2009},
\citet{Bowey2021} and \citet{BH2022} included them in three-component
models of Sakurai's Object and I include them here\footnote{Overtone
features due to other materials did not match the Mon~R2~IRS~3
spectrum \citep{BH2005}, with the exception of the 'melilite' spectrum
which was later found to be contaminated with carbonates
\citep{BH2022}.}.

Significantly better fits ($\chi^2_\nu$= 20.0 instead of 26.9 and
labelled {\bf BEST} in Figure~\ref{fig:natfits}(e)) were obtained by
adding the overtone spectrum of $\sim 20$--\micron-sized silicon
carbide grains to the nCaCO$_3$ and H$_2$O model. This component
increased the power-law index slightly, reduced $\uptau_6$(H$_2$O) by
5\% and increased $\uptau_6$(--CO$_3$) by 29\%, to 0.27. In test fits
to other MW objects the bSiC component fell to zero.

\subsection{Unfitted structure in the molecular-cloud, Elias~16, sightline}
Spectra of Elias~16 and PKS~1830--211 contain weak 0.2-\micron-wide
peaks centred at 6.3~\micron\ that are not captured by the model
fits. While the PKS~1830--211 peak could be explained by noise, structure
in the Elias-16 spectrum appears to be significant. In addition, the Elias-16
6.9-\micron\ peak is narrower than the dolomite and magnesite
laboratory spectra and is blueshifted by 0.1~\micron\ in comparison to
these carbonates. 6.9~\micron-fits might be improved with spectra of
smaller ($\la 0.04 \micron)$ magnesite or dolomite grains, or low-temperature
measurements because these factors sharpen the bands. However, it is
unlikely that they can explain the wavelength shift. Since unmatched
areas below and above the dolomite fit to the 6.9~\micron\ peak in
Elias~16 are similar, I consider derived carbonate abundances and
uncertainties representative of the true values.

The two excesses at 6.3~\micron\ and 6.75~\micron\ might be explained
by a contribution from PAHs. Preliminary attempts to obtain an
upper-limit by adding a \citet{Carpentier2012} PAH spectrum were
unconstrained, but fits with the narrower bands of individual PAHs
\citep[aka][]{Mattioda2020,Chiar2021} might provide a solution.

\subsection{Summary}

H$_2$O ice is present in all these lines of sight. Every sightline,
except AFGL~989, required a carbonate component. The best models of
Mon~R2~IRS~3 included large (20~\micron-sized) SiC grains. Only
PKS~1830--211 and AFGL~989 have a substantial CH$_3$OH
component. There is unexplained excess absorption in the spectrum of
Elias~16 at 6.3~\micron\ and 6.75~\micron.

\begin{table}
\begin{minipage}{\linewidth}
  \caption{Mass, $\Sigma$,  and molecular, $m$, densities of H$_2$O ice
    evaluated from the 6--7~\micron\ spectra with published values, $m_3$, from  3.0~\micron\ spectra \citep{Gibb2004}. Uncertainties in $\Sigma$ and $m$ are  $\la 20$\% (see Section~\ref{sec:sigma}).
    \label{tab:h2oicecolden}}
\begin{tabular}{lrll}
\hline
Object             &$\Sigma$/ $10^{-6}$~gcm$^{-2}$ &$m$ / $10^{18}$cm$^{-2}$&$m_3$ / $10^{18}$cm$^{-2}$ \\
\hline  
PKS~1830--211&120&2.7&         \\
Elias~16     &110&2.4&{\it 2.5 $\pm$ 0.06}\\
S140~IRS~1   &70 &1.6&{\it 1.9 $\pm$ 0.03}\\
AFGL~989     &96 &2.1&{\it 2.4 $\pm$ 0.1}\\
Mon~R2~IRS~3 &98 &2.2&{\it 1.9 }\\
\hline
\end{tabular}
\end{minipage}
\end{table}
\section{Abundances}
\label{sec:abundances}
\subsection{H$_2$O}
\label{sec:h2oab}


Fitted H$_2$O optical depths, $\uptau_6$(H$_2$O), in Table~\ref{table}
are compared with published estimates,
$\uptau_{6.0}(B)$. \citet{Boogert2008} define
\begin{equation}
\uptau_{6.0}(B)=\uptau_6(I_3,T) + \mathrm{C}1 + \mathrm{C}2,
\end{equation}
where $\uptau_6(I_3,T)$ is the optical depth of H$_2$O ice at
6.0~\micron\ inferred from the observed depth of the
3.0~\micron\ H$_2$O band using a laboratory spectrum obtained at
temperature $T$. Components $\mathrm{C}1$ ($\lambda_{pk}=5.8$~\micron,
fwhm $\sim0.3$~\micron) and $\mathrm{C}2$ ($\lambda_{pk}=6.2$~\micron,
fwhm $\sim 0.4$~\micron) are the optical depths of two
observationally-defined profiles of unknown materials.  The
contribution of $\mathrm{C}1 + \mathrm{C}2$ to $\uptau_{6.0}(B)$ in
Elias~16, S140~IRS~1, AFGL~989 and Mon~R2~IRS~3 were 0.01, 0.02, 0.03
and 0.01, respectively.

My $\uptau_6$(H$_2$O) values are similar to $\uptau_{6.0}(B)$. Small
discrepancies (-0.04 to +0.03) are suggestive of differences in
the continua adopted, rather than the H$_2$O temperature, because
(temperature-) broadened profiles tend to reduce fitted optical
depths.

Mass and molecular densities derived from my fits are in
Table~\ref{tab:h2oicecolden} with the molecular densities, $m_3$,
derived by \citet{Gibb2004} from the 3.0~\micron\ feature of the MW
sources.  The 6.0--7.0 \micron\ values are $\pm 20$\% of $m_3$. The
molecular H$_2$O density in the PKS~1830--211 absorber is $2.7\times
10 ^{18}$~molecules cm$^{-2}$.  This is 110\% of the, molecular
density towards Elias~16, and 120--170\% the molecular density in the
MW YSOs.

\subsection{CH$_3$OH}
\label{sec:ch3ohab}
Fitted CH$_3$OH optical depths, $\uptau_6$(CH$_3$OH), are compared
with values, $\uptau_{6.85}(P)$, predicted from 3.54~\micron\ data in
Table~\ref{table} and defined below. Mass- and molecular-densities are
in Table~\ref{tab:methicecolden}.

I used the three component fits to obtain an upper limits for the
column densities in cases where three-component fits included a very
small, but finite, CH$_3$OH component but were statistically
marginally poorer than two-component H$_2$O and carbonate fits. These
sources were Elias~16 --$\uptau_6$(CH$_3$OH)=0.017 with
$\chi^2_\nu$=3.54 versus 3.52 and S140~IRS~1
$\uptau_6$(CH$_3$OH)=0.0073 with $\chi^2_\nu$ 13.055 versus 13.052.

\subsubsection{Calibration of 6.85~\micron\ CH$_3$OH abundance}
$\uptau_{6.85}(P)$ is derived from the
3.54-to-3.0-\micron\ CH$_3$OH-to-H$_2$O molecular ratios,
$R_3$(CH$_3$OH), obtained by \citet{Gibb2004}, and reproduced in
Table~\ref{tab:methicecolden}. Since the integrated band strengths of
the 6.85- and 6.0-\micron\ CH$_3$OH- and H$_2$O- bands in
Table~\ref{tab:const} are similar,
\begin{equation}
  \uptau_{6.85}(P)\approx 1.25\times\uptau_6(\mathrm{H_2O})\times
  \frac{R_3(\mathrm{CH_3OH})}{100},
\end{equation}
where the factor of 1.25 is obtained from the ratio of the fwhms of
the pseudo-Gaussian 6.0 and 6.85-\micron\ bands. 

Fitted CH$_3$OH optical depths in S140~IRS~1 ($< 0.0073$) and
Mon~R2~IRS~3 ($<10^{-6}$) are consistent with predicted values (<
0.01); Towards Elias~16 the $\uptau_6$(CH$_3$OH) upper limit is 2.8
times the predicted value (< 0.006). Fits to AFGL~989 in which
carbonates were absent, were 1.5 times the predicted value
(0.046). 

\begin{table}
\begin{minipage}{\linewidth}
  \caption{Mass and molecular densities of CH$_3$OH ice evaluated from
    the 6--7~\micron\ spectra. $R_6$(CH$_3$OH) and $R_3$(CH$_3$OH) are
    my CH$_3$OH/H$_2$O molecular ratios from the 6--7~\micron\ spectra
    and the ratios obtained by \citet{Gibb2004} from the 3--4-\micron\
    spectra, respectively. Uncertainties in $\Sigma$, $m$ and
    $R_6$(CH$_3$OH) are <20\%, <20\% and <30\%, respectively (see
    Section~\ref{sec:sigma}).
    \label{tab:methicecolden}}
\begin{tabular}{lllcc}
\hline
Object       &$\Sigma$&$m$&$R_6$(CH$_3$OH)&$R_3$(CH$_3$OH)\\
             &$10^{-6}$~gcm$^{-2}$&$10^{18}$cm$^{-2}$&\% &\%\\
\hline  
PKS~1830--211&36      &1.1     &41&\\
Elias~16\footnote{\label{foot:upper} Upper limits are based on the marginally poorer 3-component fit.}     &<5.7     &<0.18    &<7.5& <2.9\ \ \ \ \ \ \ \ \ \ \ \\
S140~IRS~1$^{\ref{foot:upper}}$   &<2.4     &<0.077   &<4.9&<7.7\ \ \ \ \ \ \ \ \ \ \ \\
AFGL~989     &24      &0.74    &35& $23 \pm 2.5$\\
Mon~R2~IRS~3 &$\sim 0$&$\sim 0$&$\sim 0$&<4.9\ \ \ \ \ \ \ \ \ \ \ \\
\hline
\end{tabular}
\end{minipage}
\end{table}


\subsubsection{Sources of uncertainty in particle abundance estimates}
Even if the optical paths of light transmitted at different
wavelengths are identical (unlikely within a YSO disc or envelope),
abundance discrepancies between estimates from different spectral
features occur due to: (i) the choice of continuum. For example, my
polynomial fit to the Elias~16 spectrum adds curvature which explains
the larger ice optical depths. (ii) the effect of very large column
densities of small grains or a smaller number of large grains which
cause the band to saturate. The effect occurs in thin film laboratory
samples \citep[see][]{HKS2003, BHK2020, BH2022} because strong bands
become rounded and eventually opaque as the film thickness (aka column
density) is increased. Hence, if the column density is high, weak
peaks will give a truer (and larger) estimate of the total abundance.


\subsection{CH$_3$OH ice in the PKS~1830--211 galaxy-absorber}
Due to the consistency of the determined 6.85~\micron\ and 3.54~\micron\ CH$_3$OH
abundances, I conclude that the high CH$_3$OH optical depth
($\uptau_6$(CH$_3$OH)=0.11) towards PKS~1830--211 is supported by the
MW fits. The galaxy-absorber $R_6$(CH$_3$OH) ratio of 41\% is similar
to the value for AFGL~989 (35\%), but much higher than ratios in the
quiescent molecular cloud towards Elias~16 (<7.5\%) and the other YSOs
(S140~IRS~1 (4.9\%) and Mon-R2~IRS~3 ($\sim 0$)). It is 3--8 times the
$R_3$(CH$_3$OH) value in MW isolated starless molecular clouds and 14
times the value in giant molecular clouds \citep[5--12\%
  ;][]{Boogert2011,Goto2021}, while the ratio in the giant Lupus and
Taurus molecular clouds and IC~5146 is \citep[<3\%;
][]{Boogert2013,Chiar2011}. The high ratio is consistent with the
detection of gas-phase CH$_3$OH absorption in the submm lensed SW
image because observations of massive star-formation regions combined
with gas-grain chemical models \citep[e.g.][]{vanderTak2000} indicate
CH$_3$OH forms on $\la 15$~Kelvin grain surfaces in molecular clouds
when HI densities $\la 10^4$cm$^{-3}$ with CO$_2$ forming
preferentially at higher densities. Evaporation occurs at $\sim 100$~K
as a consequence of star-formation. \citet{Muller2020} obtain a total
methanol gas column density $\sim 5 \times 10^{14}$ molecule cm$^{-2}$
in the SW component. Assuming the SW lensed submm component, is
responsible for the infrared absorption feature, the CH$_3$OH
solid-to-gas ratio along this line of sight is $\sim 2000$.


\subsection{Carbonates}
\begin{table}
\begin{minipage}{\linewidth}
  \caption{Mass, grain and molecular densities of carbonates. 
    AFGL~989 models do not contain carbonate dust. \label{tab:carbcolden}
~$R_6$(--CO$_3$) is the --CO$_3$/H$_2$O molecular ratio obtained from the 6--7~\micron\ spectra. Uncertainties in $\Sigma$, $n_g$, and
    $R_6$(-CO$_3$) are <20\%, <20\% <20\% and <30\%, respectively (see Section~\ref{sec:sigma}).
  }
\begin{tabular}{lllllllll}
\hline
Object&Carbonate&$\Sigma$      &$n_g$     &$m$&$R_6$(--CO$_3$)\\
&&$10^{-6}$&$10^{6}$&$10^{15}$&\\
&&gcm$^{-2}$&cm$^{-2}$&cm$^{-2}$&\%\\
\hline  
PKS~1830--211&CaCO$_3$ &1.3&350 &2.5&0.091\\
Elias~16&CaMg(CO$_3$)$_2$&3.0&310&6.3&0.26\\

S140~IRS~1& nCaCO$_3$&1.6&9300&3.1&0.19\\

Mon~R2~IRS~3& nCaCO$_3$&4.2&24000&8.1&0.37\\
\hline
\end{tabular}
\end{minipage}
\end{table}
Carbonate column densities are listed in Table~\ref{tab:carbcolden}.
Calcite (CaCO$_3$) grains, with a peak wavelength of 6.97~\micron\ and
grain size of 0.11~\micron, provided the best fit to PKS~1830--211. MW
YSOs S140~IRS1 and Mon~R2~IRS~3 were matched with smaller
($\sim 0.04$~\micron-sized) calcite grains. Elias~16 was fitted with dolomite
(CaMg(CO$_3$)$_2$), with a peak wavelength of 6.87~\micron\ and a
grain size of 0.15~\micron. AFGL~989 models do not contain carbonate dust.

The carbonate-to-H$_2$O molecular-density ratio ($R_6$(--CO$_3$))
towards PKS~1830--211 is 35\% of the value towards Elias~16 and 48 \%
of the value in S140~IRS~1, and $\sim 25$\% the ratio towards Mon
R2~IRS~3. The mass-densities towards PKS~1830--211, Elias~16, and
S140~IRS~1 are similar to those in the circumstellar environment of
Sakurai's Object (1.2--2.8)$\times 10^{-6}$gcm$^{-2}$ between 2005
April and 2008 October \citep{BH2022}. The carbonate mass-density
towards Mon~R2~IRS~3 is 1.5 times the peak value in Sakurai's Object.

\subsection{SiC in Mon~R2~IRS~3}

\begin{table}
\begin{minipage}{\linewidth}
  \caption{Mass, grain and molecular densities of bSiC in
    Mon~R2~IRS~3. ~$R_6$(SiC) is the SiC/H$_2$O molecular ratio
    obtained from the 6--7~\micron\ spectra. Uncertainties in
    $\Sigma$, $n_g$, $m$ and $R_6$(SiC) are <20\%, <20\%, <20\% and
    <30\%, respectively (see Section~\ref{sec:sigma}).
 \label{tab:sicden}}
\begin{tabular}{llll}
\hline

$\Sigma$ &$n_g $ &$m$
&$R_6(SiC)$\\ $10^{-6}$~gcm$^{-2}$&$10^{6}$~cm$^{-2}$&$10^{18}$~cm$^{-2}$&\%\\ \hline
300&0.0089\footnote{Grain number densities are not directly comparible
with those for Sakurai's Object because the samples were different;
the grain volume ratio is $\sim$ 1.95.}&1.5&68\\
\hline
\end{tabular}
\end{minipage}
\end{table}

bSiC abundances for Mon~R2~IRS~3 are given in
Table~\ref{tab:sicden}. Due to the 20~\micron-size of these grains,
the molecular and mass-densities are substantial in comparison to the
measured H$_2$O abundances: $R_6$(SiC)$\sim$68\%; the mass density is
300 $\times 10^{-6}$ gcm$^{-2}$. For comparison, bSiC mass densities
within Sakurai's Object were 54--167 $\times 10^{-6}$ gcm$^{-2}$,
or~0.18--0.6 times the inferred value towards Mon~R2~IRS~3.


\section{Ices in the 2.5--5~\micron\ PKS~1830--211 spectrum}
\label{sec:2.5-5mod}
\begin{figure}
\begin{minipage}{0.9\linewidth}
\includegraphics[bb= 20 150 400 660, width=\linewidth]{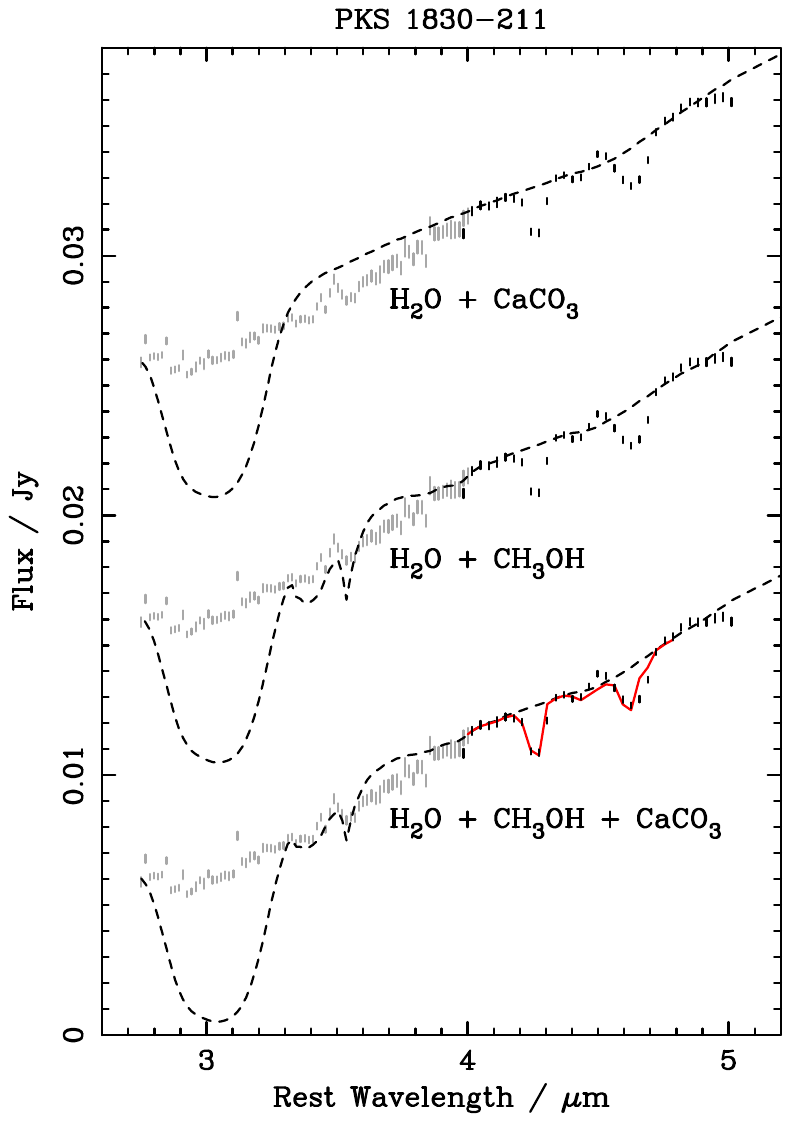}\\
\caption{Fits (solid red) to absorption features in the
  4--4.8~\micron\ SL~1 spectrum of PKS~1830--211 (black error
  bars). Dashed curves include extrapolated absorption components
  derived from the 5--8-\micron\ fits. SL~2 data (grey error bars) for
  wavelengths below 4~\micron\ were not fitted. Y-axis offsets from
  the bottom are, 0.00, 0.01 and 0.02, respectively.
\label{fig:shortfits}}
\end{minipage}
\end{figure}
\begin{figure}
\includegraphics[bb= 20 450 200 660, width=0.6\linewidth]{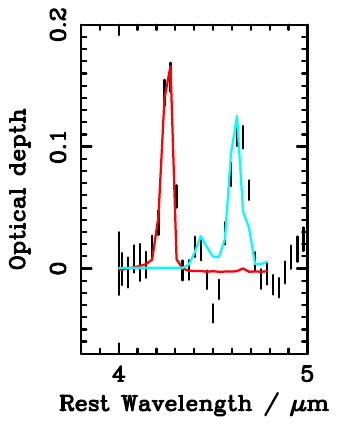}
\caption{4--5\micron\ Optical depth spectrum of PKS~1830--211 (error
  bars), with fitted components of CO:10CO$_2$ (red) and OCN$^-$ in
  irradiated formamide (cyan). The spectrum was derived with
  equation~\ref{eq:opt} using the three-component H$_2$O, CH$_3$OH and
  Calcite `continuum' in Figure~\ref{fig:shortfits}.
\label{fig:shortopt}}
\end{figure}

Figure~\ref{fig:shortfits} shows the rest-frame 2--5~\micron\ flux
spectrum of the PKS~1830--211 galaxy-absorber with laboratory spectra
scaled to optical depths obtained with the 6--7~\micron\ fits and a
modified power-law index, $\beta$ (defined, below). I attribute the
peculiar absence of a 3.0~\micron\ H$_2$O-ice band to flux
contamination from an object of unknown type in the SL~2 slit (see
Appendix~\ref{app:PKSobs}). Despite the absence of H$_2$O, the
PKS~1830--211 spectrum contains structures similar in shape and
strength to CH$_3$OH bands in the extrapolated 3-component
`continuum'. CH$_3$OH bands in the carbonate-free CH$_3$OH-rich model
are too strong.

There are clear absorption features near 4.3~$\micron$ and
4.6~\micron. I added two extra absorbers and selected the power-law
index, $\beta$ to produce the lowest $\chi^2_\nu$ values in this
wavelength range:
\begin{equation}
  F_{\nu, 4-4.8\micron}=b_o\lambda^\beta\exp(-\sum_{i=1}^5c_i\uptau_i(\lambda)).
\end{equation}
Parameters $c_1$--$c_3$ are fixed to values for H$_2$O, CH$_3$OH and
CaCO$_3$ obtained from the 5--8~\micron\ fits. The scale factor,
$b_o$ and $c_4$ and $c_5$ are fitted.

Optical depth spectra of the observations and fits, obtained with the
4--4.8~\micron\ continuum by setting $c_4$ and $c_5$ to zero, are in
Figure~\ref{fig:shortopt}. The best-matching laboratory spectra
(Table~\ref{tab:const}) were the 4.27~\micron\ CO$_2$ band in a CO :
10 CO$_2$ mixture ($\uptau = 0.17$) and the 4.61~\micron\ OCN$^-$ band
($\uptau = 0.13$) with $\beta=1.54$ and $\chi^2_\nu=7.1$ which have
been identified in MW molecular-clouds.  The fwhms of these bands are
$\sim 0.03$~\micron\ and 0.06~\micron\, similar to the spacing between
2 and 4 wavelength-intervals, respectively; this resolution is
insufficient to derive reliable abundance estimates.

\section{Summary of Results}
\label{sec:summres}
Light from blazar PKS~1830--211 passes through a face-on lensing
spiral galaxy at a redshift of z=0.886. Observations of absorption
features from IR to radio wavelengths indicate the optical path
through the galaxy includes massive star-forming regions and molecular
clouds similar to those in the Milky-Way. MW sightlines included for
comparison with the PKS~1830--211 spectrum are the quiescent
molecular-cloud towards Taurus-Elias~16, and YSOs in massive
star-forming regions (S140~1RS~1, AFGL~989, and Mon~R2~IRS~3).\\

\subsection{Carbonates and ices in star-forming regions}
I have: (i) associated the 6.9-\micron\ band with a combinations of
CH$_3$OH ice and/or carbonate dust; (ii) deciphered the carriers of
solid-state absorption features near 4.3, 4.6, 6.0 and 6.9~\micron\ in
the PKS~1830--211 absorber-rest-frame and compared the results with
those for the MW sightlines; (iii) shown CH$_3$OH : H$_2$O ratios
derived in my 6--8~\micron\ models are comparible to those obtained at
3--4~\micron\ if the observation has unsaturated 3--4~\micron\ and
6--8~\micron\ bands.

\subsection{H$_2$O, CO$_2$ and OCN$^-$}
Due to the constraint on the spectral characteristics, the range of MW
H$_2$O column densities was limited to 1.6--$2.4 \times 10^{18}$
molecules cm$^{-2}$. 6.0~\micron\ estimates of H$_2$O column density
are within $\pm 20$\% of the 3.0-\micron\ values. The H$_2$O column
density in the PKS~1830--211 galaxy-absorber is $2.7 \times 10^{18}$
molecules cm$^{-2}$.

NIR features in the PKS~1830--211 absorber were matched with the
4.27-\micron\ CO$_2$ peak in a CO : 10 CO$_2$ mixture, a
4.61-\micron\ OCN$^-$ peak but the spectral resolution is too low to
obtain estimates of their abundances.

\subsection{CH$_3$OH to H$_2$O ratios}
The PKS~1830--211 galaxy-absorber and MW YSO AFGL~989, have a
substantial CH$_3$OH:H$_2$O molecular ratios of 41\% and 35\%,
respectively. For the MW sources Elias~16, and YSOs S140~IRS~1 and
Mon~R2~IRS~3, the respective ratios are < 7.5~\%, <4.9~\% and $\sim
0$.

Whilst the MW sources have ratios within the range expected, the
PKS~1830--211 galaxy-absorber has an extremely high CH$_3$OH:H$_2$0
ratio: typical 3--4~\micron\ ratios for isolated starless
molecular-clouds within the MW are \citep[5--12\%
  ;][]{Boogert2011,Goto2021}, with values <3\% within the giant Lupus
and Taurus molecular clouds and IC~5146 is
\citep{Boogert2013,Chiar2011}). These substantial quantities of
CH$_3$OH ice appear consistent with the submm detection of gas-phase
CH$_3$OH absorption in the SW lensed component observed by
\citeauthor{Muller2021}.  Assuming the SW lensed submm component, is
responsible for the infrared absorption feature, the CH$_3$OH
solid-to-gas ratio along this line of sight is $\sim 2000$.


\subsection{Carbonates in MW molecular-clouds, YSOs and the PKS~1830--211 galaxy absorber}
\begin{figure}
  \includegraphics[bb=10 85 495
    500,width=\linewidth,clip=]{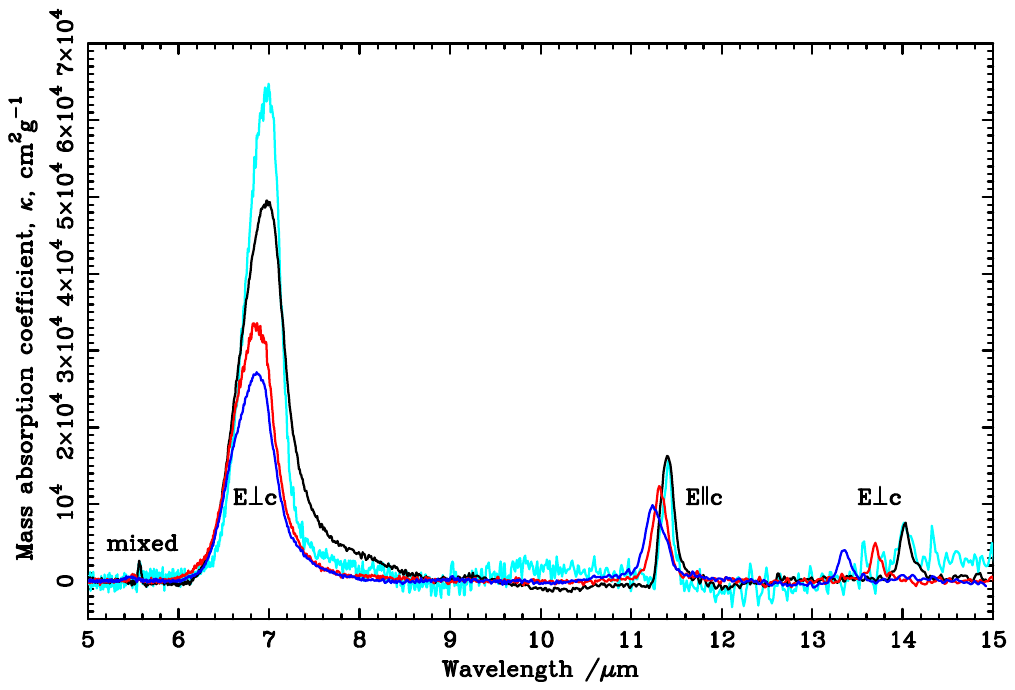}
\caption{Wavelength comparison of 0.11-\micron\ calcite
  (black),\ $\sim$0.04-\micron\ calcite
  (cyan),\ 0.15-\micron\ dolomite (red),\ 0.15-\micron\ magnesite
  (blue) in data from \citet{BH2022}. }
\label{fig:carbspec}
\end{figure}

With the exception of AFGL~989, fits to the 6.9~\micron-band of every
sightline required a contribution from carbonate dust. The best fit to
PKS~1830--211 was obtained with calcite (CaCO$_3$) grains, with a peak
wavelength of 6.97~\micron\ and grain size of 0.11~\micron. S140~IRS1
and Mon~R2~IRS~3 were matched with smaller (0.04~\micron-sized)
calcite grains. Elias~16 was fitted with dolomite CaMg(CO$_3$)$_2$,
with a peak wavelength of 6.87~\micron\ and a grain size of
0.15~\micron. The carbonate-to-H$_2$O molecular-density ratio towards
PKS~1830--211 is 39~\% of the value towards Elias~16, 50~\% of the
value towards S140~IRS~1, and $\sim 24$\% of the ratio towards Mon
R2~IRS~3.

\subsection{20-~\micron-sized SiC grains in Mon~R2~IRS~3?}
The best models of Mon~R2~IRS~3 included large (20~\micron-sized) SiC
grains to fit a plateau at 6.4~\micron.  Modelled molecular densities
of 20~\micron-sized SiC grains in Mon~R2~IRS~3 are 68~\% of the
measured H$_2$O abundances; the mass density is 300 $\times 10^{-6}$
gcm$^{-2}$. For comparison, bSiC mass densities within Sakurai's
Object were 54--167 $\times 10^{-6}$ gcm$^{-2}$, or~0.18--0.6 times
the inferred value towards Mon~R2~IRS~3.

\section{Discussion: Impact of astronomical carbonate dust}
\label{sec:impact}
\subsection{Additional spectral features}
\begin{figure}
\includegraphics[bb= 20 450 200 660, width=0.6\linewidth]{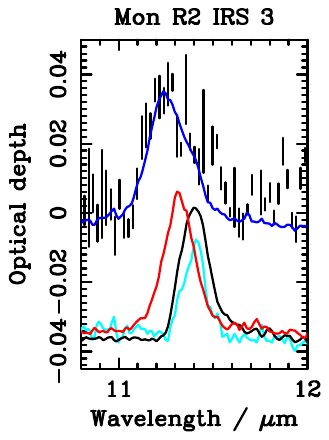}

\caption{Optical depth spectrum derived from the 11.2-\micron\ excess
  absorption in Mon~R2~IRS~3 (error bars) identified by
  \citet{Bregman2000} which is well-matched by the 0.15~\micron-thick
  magnesite optical depth spectrum (MgCO$_3$; blue). 0.15~\micron\ dolomite (red), and 0.11~$\mu$m and $\sim 0.04$~$\micron$ calcite spectra (black and cyan, respectively) are offset in the y-axis by -0.03 for clarity.}.
\label{fig:bregopt}
\end{figure}

Since carbonates contribute to the astronomical 6.9-\micron\ band
weaker bands near 5.6~\micron, 11.3 ~\micron\ and
13.2--14~\micron\ might also be detected (see
Figure~\ref{fig:carbspec}). However, due to the very large difference
in band strength, the occurrance of both sets of features in a single
sightline is likely to be infrequent. This extreme difference in band
strength was the reason a very small carbonate fraction (<0.1\%) was
missed by \citet{BH2005} in their melilite overtones study. Grains
with strong 5.6~\micron, 11.3 ~\micron\ and 13.2--14~\micron\ peaks
are very likely to be opaque at 6.9~\micron. Under controlled
laboratory conditions this property of particulate samples is a useful
method for band-strength calibration, provided it is possible to
measure the thickness of extremely thin films. After appropriate
calibration, this property might become useful in the interpretation
of astronomical data.

The effect of cation substitution of Ca$^{2+}$ with Mg$^{2+}$
relatively difficult to discern within the broad 6.9~\micron\ band,
but is much more visible in the narrow longer
wavelength-bands. \citet{Bregman2000} identified excess absorption
11.3~\micron\ in the ratioed spectrum of Mon~R2~IRS~3 and identified
it with PAHs due to its resemblance to emission bands in planetary
nebulate, HII regions and the ISM. I took the natural logarthm of
their ratio to obtain the optical depth spectrum in
Figure~\ref{fig:bregopt}; it closely resembles the 11.25~\micron\ band
in magnesite. Scaling the laboratory spectrum by eye, gives a match to
the 11.3~\micron\ optical depth, when $\uptau_{6.9}=0.11$ and mass-
and molecular- densities of $4.1\times 10^{-6}$ gcm$^{-2}$ and
$8.6\times 10^{15}$, respectively. These values are $\pm 5$\% of the
densities evaluated using nCaCO$_3$ in the 6.9~\micron\ fits of the
Mon~R2~IRS~3 ISO data which is suggestive of a match to carbonates
with the caveat that the H$_2$O+ MgCO$_3$ fit to the
5--8~\micron\ spectrum was poorer ($\chi^2_\nu \sim 70$) than that for
H$_2$O and nCaCO$_3$ ($\chi^2_\nu \sim 27$), or CH$_3$OH ($\chi^2_\nu
\sim 42$) due to mis-matches in peak wavelength and feature width.

\subsection{Grain orientation and potential for polarization effects}
Carbonate crystals are optically anisotropic: the 6.9~\micron\ and
13.2--14-\micron\ peaks are sensitive to $\boldmath{E} \perp
\boldmath{c}$ while the 11.3~\micron\ peak is responsive to
$\boldmath{E} \parallel \boldmath{c}$.  Hence, if the optical path is
the same for the 6.9~\micron\ and 11.3~\micron\ bands
similar measurements of column density at both wavelengths is
indicative of random crystal orientation. In astronomical
observations, e.g. towards Mon~R2~IRS~3, this might indicate an
insensitivity of carbonate dust to the local magnetic field and an
absence of Fe in these grains.  Fe-bearing carbonates were not
measured by \citet{BH2022} because the strongest peak in the
Fe-carbonate end-member (siderite, FeCO$_3$) is at $\sim
7.03$~\micron\ and longward of the astronomical bands, but clearly
more observations and laboratory data are needed to confirm these
hypotheses.

Carbonate dust may also provide an explanation for a narrow
polarization feature at 11.3~\micron\ in the N-band polarization
spectrum of AFGL~2591 which was tentatively attributed to an annealed
(i.e. crystalline) silicate component \citep{Aitken1988} before the
infrared-space-observatory revealed that crystalline silicates were
reasonably common in circumstellar environments. The feature is known
to be persistent over time due to the use of the source as a
position-angle standard (170$^\circ$) \citep{Smith2000}. Carbonate
orientation might be due to a physical association with flowing gas
rather than magnetic fields because Fe-free carbonates would be
insensitive to the magnetic field unless they contained magnetic
inclusions.


\subsection{Impact on chemical depletion measurements}
Observations of abundances in the transitions between the diffuse
medium (atomic hydrogen column density, $N_H <10^{21}$ cm$^{-2}$;
visual extinction, A$_V$<1), the translucent medium ($10^{21} < N_H <
10^{22}$ cm$^{-2}$; A$_V$ 3--5) and dark molecular clouds ($N_H
>10^{22}$cm$^{-2}$; $A_V>5$) established the occurrance of a rapid
removal of 30--50\% of the available oxygen atoms from the
interstellar gas as the density increases
\citep{Jenkins2009,Whittet2010}. \citet{Jones2019} argued for the
presence of (possibly undetectible) cyclic organic carbonates (COCs),
or other carrier with a C:O ratio of 1:3 which mimimises the requred
carbon depletion into an O-rich phase, but could find no observational
studies of COC spectroscopic signatures which might occur at
5.5--5.8~\micron\ and 7.8--8.2~\micron\ \footnote{These bands do not
appear in the 6--7~\micron-features in the circumstellar environment
of Sakurai's Object either, \cite{Bowey2021}}. The 6.9~\micron-band
of inorganic carbonates were mentioned in a footnote, presumably
because they had been ruled out of contention by \citet{Keane2001} or
due to the absence of laboratory data for them. My analysis of the
6.9~\micron\ feature in molecular cloud environments and
identification of carbonate dust seems to seems to fulfill the
abundance constraints, especially since the authors allowed a minor
depletion ($<20\%$) of other metals including Mg onto the grains in
these transition regions.

\section{Conclusion}
\label{sec:conclusions}
Solid-state infrared absorption features within dense molecular clouds
are useful tracers of the physical and chemical conditions within the
Milky-Way and other galaxies because the dust is a repository of
information about stellar evolution and metallicity, and is the source
material for new planets. IR spectra are used to determine the
mineralogy of oxygen-rich refractory materials because these materials
do not have characteristic spectral features in radio and submm bands
and their optical bands are obscured by the high visual extinction and
stronger atomic and molecular lines. The analysis has required data
from several laboratories which specialise in chemical synthesis,
infrared spectroscopy of ices and organics, and the Earth and
planetary sciences, as well as reinterpretation of astronomical
observations with simple empirical models
($\chi^2$-fitting).\\

I have interpreted spectral features near 6.0 and 6.9~\micron\ which
appear in molecular clouds and YSOs within the MW and the $z=0.886$
galaxy absorber in the line of sight to PKS~1830--211.  To determine
the proportions of three-component models, four MW sources were
selected using two criteria: (i) the peaks must be curved (not
flattened) because distorted spectral shapes indicate grain densities
or grain sizes which are too large for light transmission, and (ii)
there must be published H$_2$O to CH$_3$OH ice ratios derived from 3.0
to 4.0\micron\ observations to calibrate the 5 to
8-\micron\ fits. These constraints limit the range of explored MW
H$_2$O-ice column densities to 1.6--$2.4 \times 10^{18}$
molecules~cm$^{-2}$; the H$_2$O ice column density in the galaxy
absorber is $2.7\times 10^{18}$ molecules cm$^{-2}$ with an
uncertainty of $\pm 10$ to $\pm 20$ \%.

Uncertainties in molecular ratios are estimated to be $\sim \pm 30$ \%
of the quoted values. Evaluations of CH$_3$OH : H$_2$O ratio in the
ices from 6--8~\micron-spectra of PKS~1830--211 and AFGL~989 are high
at 41\% and 35\%, respectively. For Elias~16, S140~IRS~1 and
Mon~R2~IRS~3 the respective ratios are < 7.5\%, 4.9\% and $\sim 0$.

Every sightline, except AFGL~989, required a carbonate component with
grain-sizes in the 0.04--0.15~\micron-range. PKS~1830--211,
S140~IRS~1, and Mon~R2~IRS~3 spectra were all matched with calcite
(CaCO$_3$). The molecular-cloud sightline towards Elias~16 was better
matched with 0.15~\micron-sized dolomites (CaMg(CO$_3$)$_2$. However,
the carbonate fit to Elias~16 might be improved by using narrower
laboratory spectra produced by small ($\sim 0.04$~\micron-sized)
and/or low-temperature ($\sim 10$~K) Mg-bearing carbonates which were
unavailable. Unexplained excesses in the spectrum of Elias~16 at
6.3~\micron\ and 6.75~\micron\ could potentially be associated with
PAHs.

The Mon~R2~IRS~3 spectrum may indicate a population of much larger
(20~\micron-sized) SiC grains which contribute to a plateau near
6.4~\micron. Due to their large size, the SiC to H$_2$O molecular
ratio is 68 \%.

Observations with NIRSpec and MIRI on JWST and future far-infrared
instruments could enhance our understanding of the dust mineralogy and
improve the link between meteoritics, planetary science and
astrophysics of systems beyond the Solar System as far as
high-redshift galaxies. IR spectra of solid-state features in
satistically significant samples of high redshift galaxies could aid
studies of star-formation rate, metallicity, and physical conditions
at extended look-back times if there are suitable background sources.
Laboratory experiments are required to determine mechanisms for
carbonate formation in astronomical environments, and to reduce
systematic uncertainties in the abundance of dust species in dense
environments.



\section*{Acknowledgements}
The author was funded by a 2-yr Science and Technology Research
Council Ernest Rutherford Returner Fellowship (ST/S004106/1) plus a 6
month extension from Cardiff University; an additional 12 months of
her time was unremunerated. She would like to thank A. M. Hofmeister
for collaborating to produce the carbonate laboratory data, and the
anonymous referees for their time, diligence and their request that
she consider the broader implications of the work. LR Spitzer spectra
were obtained from the NASA/IPAC Infrared Science Archive IRSA, which
is funded by the National Aeronautics and Space Administration and
operated by the California Institute of Technology, and the Combined
Atlas of Sources with Spitzer/IRS Spectra (CASSIS), a product of the
Infrared Science Center at Cornell University, supported by NASA and
JPL. Observations were made with the Spitzer Space Telescope, which
was operated by the Jet Propulsion Laboratory, California Institute of
Technology under a contract with NASA.

\section*{Data Availability}
Carbonate data published by ~\citet{BH2022} are available from
https://zenodo.org/communities/mineralspectra/. Other data underlying
this article will be shared on reasonable request to the corresponding
author.









\appendix
\section{PKS~1830--211 Spitzer observations}
\label{app:PKSobs}

\begin{figure}
\begin{minipage}{\linewidth}
\includegraphics[bb=20 300 500 630, width=\linewidth]{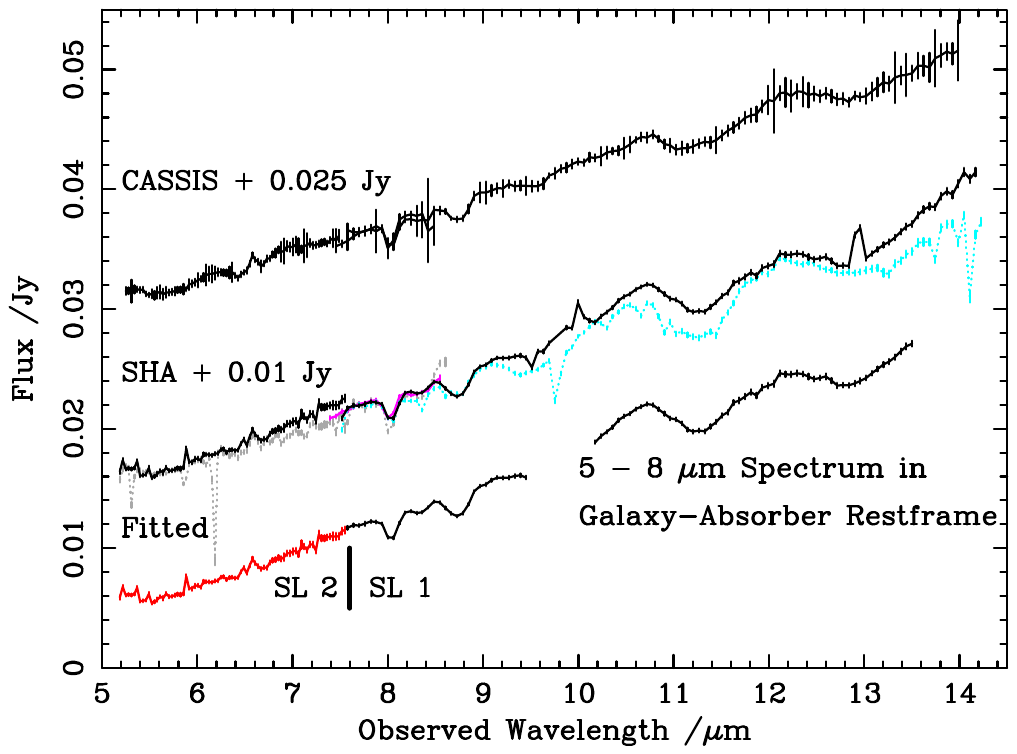}
\caption{Selection of PKS~1830--211 Spitzer data for modelling. SL~1
  and SL~2 data from the Spitzer Heritage Archive Data for beams {\bf
    005} and {\bf 003} (black solid curves) were used to produce the
  observations modelled in the paper. {\bf 003} data which overlap
  SL~1 {\bf 005} (solid magenta), and the grey and cyan dotted curves for beams
  {\bf 002} and {\bf 004}, respectively, were discarded. The red
  portion of the `Fitted Data' indicates the SL-2 spectrum contaminated by an
  extra source and the bar is at join between the SL~2 and SL~1 spectra. \label{fig:pksflx}}
\end{minipage}
\end{figure}

PKS~1830--211 was observed with the Infrared Spectrometer (IRS)
\citep{Houck2004} on the Spitzer Space Telescope \citep{Werner2004} on
2008 October 05 as part of programme 50783 (PI Kulkarni, V. P.);
Astronomical Observation Request (AOR) 26905856 was originally
reduced, published and interpreted by~\citet{Aller2012}.

Data from the Combined Atlas of Sources with Spitzer/IRS Spectra
(CASSIS) Archive \citep{Leb2011} from pipeline S18.18.0 and Level~2
data from the Spitzer Heritage Archive (SHA) retrieved in 2020 are
compared in Figure~\ref{fig:pksflx}. Due to the redshift of the
absorber ($z=0.886$), the 4--5~\micron\ and 6-7~\micron\ features of
interest were observed by order Short-Low (SL)~1 at wavelengths of
8--9~\micron\ and 10--14~$\mu$m, respectively.

\subsection{Data selection}
IRS observations were obtained by nodding between two beams; SL~1
beams in the SHA spectrum in Figure~\ref{fig:pksflx} are denoted {\bf
  004} (dotted cyan) and {\bf 005} (solid black) in the SHA
archive. If the source had been perfectly centred on the slit spectra
from the two beams would be nearly identical. However, beyond
9~\micron\ the {\bf 004}-fluxes are lower and noisier than the {\bf
  005} data indicating a loss of signal. Since the CASSIS pipeline
combines the {\bf 004} and {\bf 005} beams, the long-wavelength part
of the spectrum is 6\% lower than {\bf 005} and there is reduced
contrast in the spectral features. Hence, I discarded {\bf 004} data
except for using it to justify an interpolation across the
13-\micron\ bad-pixel spike in {\bf 005}. The result was trimmed to
the 10.2--13.5$\mu$m range to exclude a spectral curvature due to an
artefact known as the 14-$\mu$m teardrop \citep[see][]{IRShandbook}
and blue-shifted to the rest-frame of the galaxy-absorber. Error bars
are the root-mean-square uncertainties in {\bf 005}. The SL~2 spectrum
from the {\bf 003} beam (red) were scaled by a factor of 0.92 to match
SL~1 and trimmed at 7.6~\micron. Data from SL~2 {\bf 002} (grey) were
discarded due to the larger number of bad points.

\subsection{Contamination of the SL~2 slit by an unknown source}
Order SL~2 observations covered 3.0--5~\micron\ bands in the
absorber rest-frame. Unfortunately, these data are contaminated by the
presence of an additional object of unknown type in the SL~2 slit.
The source, at J2000 coordinates 18:33:40.444 -21:04:35.24, is
variously catalogued as a star and a galaxy; it increases in
brightness from 17.4~mag in the B band to 14.372~mag at K in the NOMAD
catalogue \citep{Zacharias2005} but does not seem to affect spectral
shapes in the (magenta) region of SL~2 which overlaps SL~1.

\section{Calculation of Column Densities}
\label{app:abuncalc}
\subsection{Mass Density and Mass Absorption Coefficient}
\label{app:mac}
The mass column density, $\Sigma_i$, of each component is,
\begin{equation}
\Sigma_i=\frac{c_i}{\kappa_{pk}}
\label{eq:massden}
\end{equation}
where $\kappa_{pk}$ is the mass absorption coefficient of the appropriate laboratory spectrum at the peak
wavelength of the absorber,
$\kappa_{pk}$ is given by,
\begin{equation}
\kappa_{pk}=\frac{10^4 T_{pk}}{\rho d},
\label{eq:kappa}
\end{equation}
where $T_{pk}$ is the optical depth of the sample of thickness $d$ at
wavelength $\lambda_{pk}$.

\subsection{Grain Number Density}
\label{app:grainden}
The grain number density, $n_g$ of each component is
\begin{equation}
  n_g=\frac{\Sigma_i}{\rho d^3}
\end{equation}  
where $\rho d^3$, is the mass of a single grain. This mass calculation
includes three approximations: (i) that the mass densities of
terrestrial materials are similar to those of interstellar materials,
(ii) that the thickness of a powder film thin can be used as a
representative grain size and (iii) that the geometry is approximately
cubic.  Film thickness is assumed to be representative of the largest
grain sizes in the sample because boundary reflections between grains
smaller than the thickness of a compressed powder are minimized due to
the absence of airspaces (or matrix material) with a significantly
different refractive index.  However, the powder-measurement will
represent the average of a range of crystal orientations. Cubic grains
are isotropic so the orientations will be truly random. Problems arise
from elongated grains which will preferentially lie with their long
axes perpendicular to the compression axis; film thickness is
difficult to measure in very small samples (e.g. the nCaCO$_3$ sample)
and air spaces will be imperfectly removed. However, my assumption is
no worse than the ubiquitous use of astronomical assumption of
spheroidal or ellipsoidal shapes. By definition crystals have specific
angular shapes, e.g. some forms of SiC are cubic, carbonates are
rhombohedral. Grinding mineral samples will knock some corners
off. Care is taken to avoid this as far as possible
\citep[e.g.see][]{BHK2020} because over-grinding destroys the crystal
structure and alters the infrared spectrum
\citep[e.g.][]{Farmer1958,Imai2009}.


\subsection{Number of molecular absorbers and integrated band strength}
\label{app:molabs}
The number of absorbers, $m_i$, is obtained from the integrated band
strength, $A_i$, (cm molecule$^{-1}$). The integrated band strength of a laboratory sample
 is given by
\begin{equation}
  A_i=\frac{10^4 m_r}{Zd} \int^{\nu_2}_{\nu_1} T_i(\nu)d\nu,
  \label{eq:sigmaint}
\end{equation}
where and $\nu_1$ and $\nu_2$ are the frequencies at the edges of the
band, $m_r$ is the relative molecular mass of the sample and $Z$ is
Avogadro's number. The optical depth, $T(\nu)$, at frequency $\nu$,
has a sample-thickness $d$ in \micron. For a Gaussian peak the
integral can be approximated by $\Delta_\nu T_{pk}$ where,
$\Delta_\nu$ is the full-width-half-maximum in wavenumbers (cm$^-1$)
and $T_{pk}$ is the peak optical depth. The number of molecular
absorbers is given by
\begin{equation}
  m_i=c_i \Delta_\nu/A_i.
\label{eq:mol}
\end{equation}

It can also be shown, by ratioing equations \ref{eq:kappa}
and \ref{eq:sigmaint}, that for gaussian peaks,
\begin{equation}
  A_i \approx \frac{m_r\Delta_\nu\rho}{Z} \kappa_{pk}.
\end{equation}
and that $m_i$ and $\Sigma$ are equivalent, but non-identical measures
of abundance due to the different calculation methods.


\bsp	
\label{lastpage}

\begin{thebibliography}{99}
\bibitem[\protect\citeauthoryear{Aitken et al.}{1988}]{Aitken1988} Aitken D.~K., Roche P.~F., Smith C.~H., James S.~D., Hough J.~H., 1988, MNRAS, 230, 629. doi:10.1093/mnras/230.4.629
\bibitem[\protect\citeauthoryear{Allamandola et al.}{1992}]{allamandola1992} Allamandola L.~J., Sandford S.~A., Tielens A.~G.~G.~M., Herbst T.~M., 1992, ApJ, 399, 134. doi:10.1086/171909
\bibitem[\protect\citeauthoryear{Allen}{1972}]{Allen1972} Allen D.~A., 1972, ApJL, 172, L55. doi:10.1086/180890
\bibitem[\protect\citeauthoryear{Aller \& Kulkarni}{2021}]{AllerJWST2021} Aller M.~C., Kulkarni V.~P., 2021, jwst.prop, 2441
\bibitem[\protect\citeauthoryear{Aller et al.}{2012}]{Aller2012} Aller M.~C., Kulkarni V.~P., York D.~G., Vladilo G., Welty D.~E., Som D., 2012, ApJ, 748, 19. doi:10.1088/0004-637X/748/1/19
\bibitem[\protect\citeauthoryear{van Broekhuizen, Keane, \& Schutte}{2004}]{vanBroekhuizen2004} van Broekhuizen F.~A., Keane J.~V., Schutte W.~A., 2004, A\&A, 415, 425. doi:10.1051/0004-6361:20034161
\bibitem[\protect\citeauthoryear{van Broekhuizen et al.}{2006}]{vanBroekhuizen2006} van Broekhuizen F.~A., Groot I.~M.~N., Fraser H.~J., van Dishoeck E.~F., Schlemmer S., 2006, A\&A, 451, 723. doi:10.1051/0004-6361:20052942
  
\bibitem[\protect\citeauthoryear{Boogert et al.}{2008}]{Boogert2008} Boogert A.~C.~A., Pontoppidan K.~M., Knez C., Lahuis F., Kessler-Silacci J., van Dishoeck E.~F., Blake G.~A., et al., 2008, \apj, 678, 985. doi:10.1086/533425

\bibitem[\protect\citeauthoryear{Boogert et al.}{2011}]{Boogert2011} Boogert A.~C.~A., Huard T.~L., Cook A.~M., Chiar J.~E., Knez C., Decin L., Blake G.~A., et al., 2011, \apj, 729, 92. doi:10.1088/0004-637X/729/2/92  

\bibitem[\protect\citeauthoryear{Boogert et al.}{2013}]{Boogert2013} Boogert A.~C.~A., Chiar J.~E., Knez C., {\"O}berg K.~I., Mundy L.~G., Pendleton Y.~J., Tielens A.~G.~G.~M., et al., 2013, ApJ, 777, 73. doi:10.1088/0004-637X/777/1/73

  

\bibitem[\protect\citeauthoryear{Bowey \& Hofmeister}{2005}]{BH2005}Bowey, J. E.,Hofmeister, A. M. 2005,  MNRAS, 358, 1383 

\bibitem[\protect\citeauthoryear{Bowey, Hofmeister, \& Keppel}{2020}]{BHK2020} Bowey J.~E., Hofmeister A.~M., Keppel E., 2020, MNRAS, 497, 3658. doi:10.1093/mnras/staa2227  

\bibitem[\protect\citeauthoryear{Bowey}{2021}]{Bowey2021} Bowey J.~E., 2021, MNRAS, 505, 568. doi:10.1093/mnras/stab1305  

\bibitem[\protect\citeauthoryear{Bowey \& Hofmeister}{2022}]{BH2022}Bowey J.~E., Hofmeister A.~M., 2022, MNRAS, 513, 1774. doi:10.1093/mnras/stac993



\bibitem[\protect\citeauthoryear{Bowey, Adamson, \& Whittet}{1998}]{Bowey1998} Bowey J.~E., Adamson A.~J., Whittet D.~C.~B., 1998, MNRAS, 298, 131. doi:10.1046/j.1365-8711.1998.01640.x

\bibitem[\protect\citeauthoryear{Bregman, Hayward, \& Sloan}{2000}]{Bregman2000} Bregman J.~D., Hayward T.~L., Sloan G.~C., 2000, ApJL, 544, L75. doi:10.1086/317294

  
\bibitem[\protect\citeauthoryear{Brucato, Baratta, \& Strazzulla}{2006}]{Brucato2006} Brucato J.~R., Baratta G.~A., Strazzulla G., 2006, A\&A, 455, 395. doi:10.1051/0004-6361:20065095

\bibitem[\protect\citeauthoryear{Carpentier et al.}{2012}]{Carpentier2012} Carpentier Y., F{\'e}raud G., Dartois E., Brunetto R., Charon E., Cao A.-T., d'Hendecourt L., et al., 2012, \aap, 548, A40. doi:10.1051/0004-6361/201118700

\bibitem[\protect\citeauthoryear{Ceccarelli et al.}{2002}]{Ceccarelli2002} Ceccarelli C., Caux E., Tielens A.~G.~G.~M., Kemper F., Waters L.~B.~F.~M., Phillips T., 2002, A\&A, 395, L29. doi:10.1051/0004-6361:20021490

\bibitem[\protect\citeauthoryear{Chiar et al.}{2000}]{Chiar2000} Chiar J.~E., Tielens A.~G.~G.~M., Whittet D.~C.~B., Schutte W.~A., Boogert A.~C.~A., Lutz D., van Dishoeck E.~F., et al., 2000, ApJ, 537, 749. doi:10.1086/309047

\bibitem[\protect\citeauthoryear{Chiar et al.}{2007}]{Chiar2007} Chiar J.~E., Ennico K., Pendleton Y.~J., Boogert A.~C.~A., Greene T., Knez C., Lada C., et al., 2007, ApJL, 666, L73. doi:10.1086/521789

\bibitem[\protect\citeauthoryear{Chiar et al.}{2011}]{Chiar2011} Chiar J.~E., Pendleton Y.~J., Allamandola L.~J., Boogert A.~C.~A., Ennico K., Greene T.~P., Geballe T.~R., et al., 2011, ApJ, 731, 9. doi:10.1088/0004-637X/731/1/9
  
\bibitem[\protect\citeauthoryear{Chiar et al.}{2021}]{Chiar2021} Chiar J.~E., de Barros A.~L.~F., Mattioda A.~L., Ricca A., 2021, ApJ, 908, 239. doi:10.3847/1538-4357/abd6e8




\bibitem[\protect\citeauthoryear{Day et al.}{2013}]{Day2013} Day S.~J., Thompson S.~P., Parker J.~E., Evans A., 2013, A\&A, 553, A68. doi:10.1051/0004-6361/201321138


  \bibitem[\protect\citeauthoryear{de Graauw et al.}{1996}]{deGraauw1996} de Graauw T., Haser L.~N., Beintema D.~A., Roelfsema P.~R., van Agthoven H., Barl L., Bauer O.~H., et al., 1996, A\&A, 315, L49

\bibitem[\protect\citeauthoryear{Dohn{\'a}lek et al.}{2003}]{Dohnalek2003} Dohn{\'a}lek Z., Kimmel G.~A., Ayotte P., Smith R.~S., Kay B.~D., 2003, JChPh, 118, 364. doi:10.1063/1.1525805
  
\bibitem[\protect\citeauthoryear{Evans et al.}{2020}]{Evans2020} Evans A., Gehrz R.~D., Woodward C.~E., Banerjee D.~P.~K., Geballe T.~R., Clayton G.~C., Sarre P.~J., et al., 2020, \mnras, 493, 1277. doi:10.1093/mnras/staa343

\bibitem[\protect\citeauthoryear{Evans et al.}{1989}]{Evans1989} Evans N.~J., Mundy L.~G., Kutner M.~L., Depoy D.~L., 1989, ApJ, 346, 212. doi:10.1086/168002

\bibitem[\protect\citeauthoryear{Farmer}{1958}]{Farmer1958} Farmer V.~C., 1958, MinM, 31, 829. doi:10.1180/minmag.1958.031.241.03

\bibitem[\protect\citeauthoryear{Galván-Ruiz et al.}{2009}]{GR2009}Galván-Ruiz M, Hernández, J., Baños L., Noriega-Montes J., Rodríguez-García M. E. 2009 Journal of Materials in Civil Engineering, 21, 694, DOI: 10.1061/(ASCE)0899-1561(2009)21:11(694)

\bibitem[\protect\citeauthoryear{Gerakines et al.}{1995}]{Gerakines1995} Gerakines P.~A., Schutte W.~A., Greenberg J.~M., van Dishoeck E.~F., 1995, A\&A, 296, 810

\bibitem[\protect\citeauthoryear{Gibb \& Whittet}{2002}]{GibbWhittet2002} Gibb E.~L., Whittet D.~C.~B., 2002, ApJL, 566, L113. doi:10.1086/339633
  
\bibitem[\protect\citeauthoryear{Gibb et al.}{2004}]{Gibb2004} Gibb E.~L., Whittet D.~C.~B., Boogert A.~C.~A., Tielens A.~G.~G.~M., 2004, ApJS, 151, 35. doi:10.1086/381182

\bibitem[\protect\citeauthoryear{Goto et al.}{2021}]{Goto2021} Goto M., Vasyunin A.~I., Giuliano B.~M., Jim{\'e}nez-Serra I., Caselli P., Rom{\'a}n-Z{\'u}{\~n}iga C.~G., Alves J., 2021, A\&A, 651, A53. doi:10.1051/0004-6361/201936385

 \bibitem[\protect\citeauthoryear{Grishko \& Duley}{2002}]{Grishko2002} Grishko V.~I., Duley W.~W., 2002, \apj, 568, 448. doi:10.1086/338926

 \bibitem[\protect\citeauthoryear{Grellmann et al.}{2011}]{Grellmann2011} Grellmann R., Ratzka T., Kraus S., Linz H., Preibisch T., Weigelt G., 2011, A\&A, 532, A109. doi:10.1051/0004-6361/201116699

\bibitem[\protect\citeauthoryear{Hensley \& Draine}{2020}]{Hensley2020} Hensley B.~S., Draine B.~T., 2020, ApJ, 895, 38. doi:10.3847/1538-4357/ab8cc3

\bibitem[\protect\citeauthoryear{Hofmeister, Keppel \& Speck}{2003}]{HKS2003} Hofmeister, A. M., Keppel, E., Speck, A. K., 2003, \mnras, 345, 16

\bibitem[\protect\citeauthoryear{Hofmeister et al.}{2009}]{Hof2009} Hofmeister A.~M., Pitman K.~M., Goncharov A.~F., Speck A.~K., 2009, \apj, 696, 1502. doi:10.1088/0004-637X/696/2/1502

\bibitem[\protect\citeauthoryear{Hoppe et al.}{1994}]{Hoppe1994} Hoppe P., Amari S., Zinner E., Ireland T., Lewis R.~S., 1994, ApJ, 430, 870. doi:10.1086/174458

\bibitem[\protect\citeauthoryear{Hudgins et al.}{1993}]{Hudgins1993} Hudgins D.~M., Sandford S.~A., Allamandola L.~J., Tielens A.~G.~G.~M., 1993, ApJS, 86, 713. doi:10.1086/191796

\bibitem[\protect\citeauthoryear{Houck et al.}{2004}]{Houck2004} Houck J.~R., Roellig T.~L., van Cleve J., Forrest W.~J., Herter T., Lawrence C.~R., Matthews K., et al., 2004, \apjs, 154, 18. doi:10.1086/423134

\bibitem[\protect\citeauthoryear{Hough et al.}{2008}]{Hough2008} Hough J.~H., Aitken D.~K., Whittet D.~C.~B., Adamson A.~J., Chrysostomou A., 2008, MNRAS, 387, 797. doi:10.1111/j.1365-2966.2008.13274.x

\bibitem[\protect\citeauthoryear{Imai, et al.}{2009}]{Imai2009} Imai Y., Koike C., Chihara H., Murata K., Aoki T., Tsuchiyama A., 2009, \aap, 507, 277
\bibitem[\protect\citeauthoryear{IRS}{2011}]{IRShandbook} IRS Instrument Team \& Science User Support Team, Version 5.0,
S18.18 December 2011 title: IRS Instrument Handbook. https://irsa.ipac.caltech.edu/data/SPITZER/docs/irs/irsinstrumenthandbook/.

\bibitem[\protect\citeauthoryear{Jenkins}{2009}]{Jenkins2009} Jenkins E.~B., 2009, ApJ, 700, 1299. doi:10.1088/0004-637X/700/2/1299
\bibitem[\protect\citeauthoryear{Jones \& Ysard}{2019}]{Jones2019} Jones A.~P., Ysard N., 2019, A\&A, 627, A38. doi:10.1051/0004-6361/201935532
\bibitem[\protect\citeauthoryear{Keane et al.}{2001}]{Keane2001} Keane J.~V., Tielens A.~G.~G.~M., Boogert A.~C.~A., Schutte W.~A., Whittet D.~C.~B., 2001, \aap, 376, 254. doi:10.1051/0004-6361:20010936

\bibitem[\protect\citeauthoryear{Kemper et al.}{2002}]{Kemper2002} Kemper F., J{\"a}ger C., Waters L.~B.~F.~M., Henning T., Molster F.~J., Barlow M.~J., Lim T., et al., 2002, Natur, 415, 295. doi:10.1038/415295a 

\bibitem[\protect\citeauthoryear{Knez et al.}{2005}]{Knez2005} Knez C., Boogert A.~C.~A., Pontoppidan K.~M., Kessler-Silacci J., van Dishoeck E.~F., Evans N.~J., Augereau J.-C., et al., 2005, ApJL, 635, L145. doi:10.1086/499584


\bibitem[\protect\citeauthoryear{Koresko et al.}{1993}]{Koresko1993} Koresko C.~D., Beckwith S., Ghez A.~M., Matthews K., Herbst T.~M., Smith D.~A., 1993, AJ, 105, 1481. doi:10.1086/116526


  
\bibitem[\protect\citeauthoryear{Lebouteiller et al.}{2011}]{Leb2011}Lebouteiller, V., Barry, D.~J., Spoon, H.W.W., Bernard-Salas, J., Sloan, G.C., Houck, J.R., Weedman, D., 2011 \apjs, 196, 8


  
\bibitem[\protect\citeauthoryear{Lisse et al.}{2007}]{Lisse2007} Lisse C.~M., Kraemer K.~E., Nuth J.~A., Li A., Joswiak D., 2007, Icar, 187, 69. doi:10.1016/j.icarus.2006.11.019

\bibitem[\protect\citeauthoryear{Luna et al.}{2018}]{Luna2018} Luna R., Molpeceres G., Ortigoso J., Satorre M.~A., Domingo M., Mat{\'e} B., 2018, A\&A, 617, A116. doi:10.1051/0004-6361/201833463

\bibitem[\protect\citeauthoryear{Mangan et al.}{2017}]{Mangan2017} Mangan T.~P., Salzmann C.~G., Plane J.~M.~C., Murray B.~J., 2017, Icar, 294, 201. doi:10.1016/j.icarus.2017.03.012

\bibitem[\protect\citeauthoryear{Maud \& Hoare}{2013}]{MaudHoare2013} Maud L.~T., Hoare M.~G., 2013, ApJL, 779, L24. doi:10.1088/2041-8205/779/2/L24

  
\bibitem[\protect\citeauthoryear{Mattioda et al.}{2020}]{Mattioda2020} Mattioda A.~L., Hudgins D.~M., Boersma C., Bauschlicher C.~W., Ricca A., Cami J., Peeters E., et al., 2020, ApJS, 251, 22. doi:10.3847/1538-4365/abc2c8

\bibitem[\protect\citeauthoryear{McClure}{2009}]{McClure2009} McClure M., 2009, ApJL, 693, L81. doi:10.1088/0004-637X/693/2/L81

\bibitem[\protect\citeauthoryear{Muller et al.}{2020}]{Muller2020} Muller S., Jaswanth S., Horellou C., Mart{\'\i}-Vidal I., 2020, A\&A, 641, L2. doi:10.1051/0004-6361/202038978


\bibitem[\protect\citeauthoryear{Muller et al.}{2021}]{Muller2021} Muller S., Ubachs W., Menten K.~M., Henkel C., Kanekar N., 2021, A\&A, 652, A5. doi:10.1051/0004-6361/202140531


\bibitem[\protect\citeauthoryear{Preibisch \& Smith}{2002}]{PreibischSmith2002} Preibisch T., Smith M.~D., 2002, A\&A, 383, 540. doi:10.1051/0004-6361:20011772

\bibitem[\protect\citeauthoryear{Preibisch et al.}{2002}]{Preibischetal2002} Preibisch T., Balega Y.~Y., Schertl D., Weigelt G., 2002, A\&A, 392, 945. doi:10.1051/0004-6361:20021191

\bibitem[\protect\citeauthoryear{Puetter et al.}{1979}]{puetter79} Puetter R.~C., Russell R.~W., Soifer B.~T., Willner S.~P., 1979, ApJ, 228, 118. doi:10.1086/156828

\bibitem[\protect\citeauthoryear{Puetter et al.}{1977}]{puetter1977} Puetter R.~C., Russell R.~W., Soifer B.~T., Willner S.~P., 1977, BAAS

  
  
\bibitem[\protect\citeauthoryear{Rubin \& Ma}{2017}]{RubinMa2017} Rubin A.~E., Ma C., 2017, Chemie der Erde Geochemistry, 77, 325. doi:10.1016/j.chemer.2017.01.005

\bibitem[\protect\citeauthoryear{Sandford \& Walker}{1985}]{SA1985} Sandford S.~A., Walker R.~M., 1985, ApJ, 291, 838. doi:10.1086/163120

\bibitem[\protect\citeauthoryear{Smith et al.}{2000}]{Smith2000} Smith C.~H., Wright C.~M., Aitken D.~K., Roche P.~F., Hough J.~H., 2000, MNRAS, 312, 327. doi:10.1046/j.1365-8711.2000.03158.x
\bibitem[\protect\citeauthoryear{Soifer et al.}{1979}]{soifer1979} Soifer B.~T., Puetter R.~C., Russell R.~W., Willner S.~P., Harvey P.~M., Gillett F.~C., 1979, ApJL, 232, L53. doi:10.1086/183035
\bibitem[\protect\citeauthoryear{Speck, Thompson, \& Hofmeister}{2005}]{Speck2005} Speck A.~K., Thompson G.~D., Hofmeister A.~M., 2005, \apj, 634, 426. doi:10.1086/496955
\bibitem[\protect\citeauthoryear{Tercero et al.}{2020}]{Tercero2020}  Tercero B., Cernicharo J., Cuadrado S., de Vicente P., Gu{\'e}lin
  M., 2020, A\&A, 636, L7. doi:10.1051/0004-6361/202037837
\bibitem[\protect\citeauthoryear{Thompson et al.}{1998}]{Thompson1998} Thompson R.~I., Corbin M.~R., Young E., Schneider G., 1998, ApJL, 492, L177. doi:10.1086/311096
\bibitem[\protect\citeauthoryear{van der Tak, van Dishoeck, \& Caselli}{2000}]{vanderTak2000} van der Tak F.~F.~S., van Dishoeck E.~F., Caselli P., 2000, A\&A, 361, 327. doi:10.48550/arXiv.astro-ph/0008010
\bibitem[\protect\citeauthoryear{Werner et al.}{2004}]{Werner2004} Werner M.~W., Roellig T.~L., Low F.~J., Rieke G.~H., Rieke M., Hoffmann W.~F., Young E., et al., 2004, \apjs, 154, 1. doi:10.1086/422992
\bibitem[\protect\citeauthoryear{Whittet, Duley, \& Martin}{1990}]{Whittet1990} Whittet D.~C.~B., Duley W.~W., Martin P.~G., 1990, \mnras, 244, 427
\bibitem[\protect\citeauthoryear{Whittet}{2010}]{Whittet2010} Whittet D.~C.~B., 2010, ApJ, 710, 1009. doi:10.1088/0004-637X/710/2/1009
\bibitem[\protect\citeauthoryear{White}{1974}]{White1974}White, W.~B., Ch 12 in "The Infrared Spectra of Minerals", Mineralogical Society of Great Britain and Ireland 1974:01,doi:10.1180/mono-4.12 page 227.

\bibitem[\protect\citeauthoryear{Smith et al.}{2000}]{Smith2000} Smith C.~H., Wright C.~M., Aitken D.~K., Roche P.~F., Hough J.~H., 2000, MNRAS, 312, 327. doi:10.1046/j.1365-8711.2000.03158.x

\bibitem[\protect\citeauthoryear{Winn et al.}{2002}]{Winn2002} Winn, J. N., Kochanek, C. S., McLeod, B. A., et al. 2002, ApJ, 575, 103 

\bibitem[\protect\citeauthoryear{Zacharias et al.}{2005}]{Zacharias2005} Zacharias N., Monet D.~G., Levine S.~E., Urban S.~E., Gaume R., Wycoff G.~L., 2005, yCat, I/297  


\end{thebibliography}
\end{document}